\documentclass[10pt,twocolumn,letterpaper,journal]{IEEEtran}
\usepackage{siunitx}
\usepackage{balance}
\usepackage{cite}
\usepackage{epsfig}
\usepackage{times}
\usepackage{graphicx}
\usepackage[figuresright]{rotating}
\usepackage{subcaption}
\usepackage{amssymb}
\usepackage{amsmath}
\usepackage{setspace}
\usepackage{rotating}
\usepackage{multirow}
\usepackage{wrapfig}
\usepackage{booktabs}
\usepackage{array}
\usepackage{comment}
\usepackage{epstopdf}
\usepackage[utf8]{inputenc}
\usepackage{parskip}
\usepackage{amsmath}
\usepackage{indentfirst}
\usepackage{graphicx}
\graphicspath{ {images/} }
\usepackage[rightcaption]{sidecap}
\usepackage[export]{adjustbox}[2011/08/13]

\usepackage{tabu}
\usepackage{booktabs}%
\usepackage{tabularx}
\usepackage[font=footnotesize]{caption}

\usepackage[normalem]{ulem}

\let\OLDthebibliography\thebibliography
\renewcommand\thebibliography[1]{
  \OLDthebibliography{#1}
  \setlength{\parskip}{0pt}
  \setlength{\itemsep}{3pt plus 0.3ex}
}

\makeatletter
\IEEEtriggercmd{\reset@font\normalfont\fontsize{7.9pt}{8.40pt}\selectfont}
\makeatother
\IEEEtriggeratref{1}

\IEEEoverridecommandlockouts

\usepackage{algorithmic}
\usepackage{algorithm}
\floatname{algorithm}{Algorithm} \textfloatsep 0.5cm

\usepackage{colortbl}

\begin{document}
  \title{Towards Wireless Health Monitoring via Analog Signal Compression based Biosensing Platform}
\author{Xueyuan~Zhao,~\IEEEmembership{Member,~IEEE,} Vidyasagar~Sadhu,~\IEEEmembership{Student~Member,~IEEE,} Tuan~Le,~\IEEEmembership{Student~Member,~IEEE,} Dario~Pompili,~\IEEEmembership{Senior~Member,~IEEE}, Mehdi~Javanmard,~\IEEEmembership{Member,~IEEE}%
\thanks{The authors are with the Department of Electrical and Computer Engineering, Rutgers University–-New Brunswick, NJ, USA. E-mails: \{xueyuan.zhao, vidyasagar.sadhu, tuananh.le, pompili, mehdi.javanmard\}@rutgers.edu.}
\thanks{A preliminary version of this work appeared in the \emph{Proc. of the IEEE Intl. Symposium on Circuits \& Systems~(ISCAS)}, Baltimore, USA, May~2017~\cite{Zhao17}.}}%

\maketitle
\thispagestyle{empty}

\begin{abstract}
Wireless all-analog biosensor design for concurrent microfluidic and physiological signal monitoring is presented in this work. The key component is an all-analog circuit capable of compressing two analog sources into one analog signal by Analog Joint Source-Channel Coding~(AJSCC). Two circuit designs are discussed, including the stacked-Voltage Controlled Voltage Source~(VCVS) design with the fixed number of levels, and an improved design, which supports a flexible number of AJSCC levels. Experimental results are presented on the wireless biosensor prototype, composed of Printed Circuit Board~(PCB) realizations of the stacked-VCVS design. Furthermore, circuit simulation and wireless link simulation results are presented on the improved design. Results indicate that the proposed wireless biosensor is well suited for sensing two biological signals simultaneously with high accuracy, and can be applied to a wide variety of low-power and low-cost wireless continuous health monitoring applications.

\end{abstract}

\section{Introduction}\label{sec:introduction}

\textbf{Background:}
The design of wireless biosensors is essential to the realization of wireless health monitoring solutions~\cite{Soh15}. Wireless monitoring of physiological signals brings the advantages of 
portability~\cite{Rajendra17}, wearability~\cite{Yeo16}, continuous measurements~\cite{Sun16}, and improved point-of-care to patient in comparison with wired counterparts. Furthermore, microfluidic biosensors have emerged as powerful tools 
for wearable biomarkers monitoring, including electrical impedance detection~\cite{EJ12}, Giant Magnetoresistance~(GMR) 
sensing~\cite{GasterHall09}, and electrochemical detection~\cite{GaoEmaminejad16,Gholizadeh17}.
Microfluidic sensing can enable wearable sensors that are minimally invasive due to the minute sample volumes required for accurate sensing, thus promoting personalized health monitoring through continuous quantification of biomarkers. The detection of various biomolecules in blood samples using impedance-based biodetection has been demonstrated in previous works~\cite{Lin15}, which shows the potential to realize truly miniaturized wearable diagnostic platforms. Among the microfluidic sensors, impedance-based sensors typically integrate electrodes in microfluidic channels and monitor changes in impedance as particles flow over the electrodes. This class of sensors has been shown to be well suited for detecting biomarkers with high sensitivity~\cite{Mok14, Javanmard11}. Impedance-based sensors can be made portable with miniaturized analog front-end circuitry such as lock-in-amplifiers~\cite{Talukder17}, and wireless transmitters. Functionalization of microfluidic channels in impedance cytometers with different types of antibodies and enzymes can enable biomarkers detection with high specificity. 

There are various scenarios where a network of wireless body sensors monitoring both physiological signals and molecular biomarkers concentrations are beneficial to patient care. One medical scenario is monitoring hospitalized patients with cardiovascular disease. There is a vast body of literature regarding the benefits of monitoring patients with cardiovascular disease continuously. Kario et al.~\cite{Kario17} demonstrated the usage of wireless sensors to monitor the physiological signals of blood pressure and the environmental parameters for patients with cardiovascular disease. Hu et al.~\cite{Hu16} presented a microfluidic-based portable device to improve point-of-care monitoring of cardiovascular disease patients. We envision wireless wearable sensors potentially replacing wired tags used to monitor physiological signals including Electrocardiogram~(ECG), heart rate, respiratory rate, blood pressure, temperature and blood oxygen levels by electrical or optical sensors, providing continuous wireless monitoring of these vital signs. In addition, various blood biomarkers reflecting the state of the cardiovascular and endocrine system, can also be monitored in these wearable sensor networks. Potential cardiac markers of interest include levels of cardiac Troponin, BNP, creatinine, C-reactive protein~(CRP), and also complete blood cell counts. The true power of both physiological and biomarker measurements can be realized when both are combined together and the correlations between the two data types are better understood. In this work, we present a wireless biosensor design that can measure physiological signal and molecular biomarkers \textit{concurrently}. A network of body sensors monitoring various physiological parameters \textit{simultaneously} with biomarkers monitoring in blood, sweat, and exhaled breath condensate can greatly improve our understanding of disease pathogenesis. 

The circuit design of the biosensor forms the foundation of this type of hybrid wearable wireless biosensing platform. Existing wireless biosensor designs rely on digital circuits that have high complexity and are power inefficient. These limitations hinder the much needed features of large-scale deployment and battery-less monitoring. In this work, we are bridging the technological gap between hybrid sensor design and analog circuitry, by introducing a particular type of analog signal compression circuit that is aimed at compressing microfluidic and physiological signals, without the need of digital circuits. This work ultimately eliminates the operations of conversions between digital and analog signals, therefore significantly improving the power efficiency of biosensor circuits. Moreover, signal recovery performance of the proposed circuit is guaranteed for target applications.

\textbf{Existing Work:}
A sensing system was demonstrated~\cite{Imani16} where sweat-lactate concentration is measured electrochemically with a concurrent ECG measurement. Here, the sweat-lactate concentration and ECG signals are first passed to an Analog-to-Digital Converter~(ADC) for processing by a microcontroller, and then are wirelessly transmitted via Bluetooth module. Luhmann et al.~\cite{Luhmann16} demonstrated a wireless sensor that simultaneously measures the EEG and neurophysiological signal using digital transmission of the data via Bluetooth. Abrar et al.~\cite{Abrar16} presented a wireless sensor in which the sweat lactate is measured and the sensing signal is sampled with an ADC, processed by a microprocessor, and then transmitted by a digital Near-Field Communication~(NFC) wireless module. Nemiroski et al.~\cite{Nemiroski14} designed a digital mobile system for the transmission of electrochemical detection data. 
In all these systems, the major power consumption is due to the digital conversion by the ADCs and subsequent processing by the microcontrollers, which limits the battery life of the devices and the practicality of the system. Our previous work~\cite{SensorsJournal18} proposed an analog compression circuit for two signals, and introduces the technique in a generalizable context. However, it was limited to wireless circuits only and did not include biosensors for biomarker detection, which involves measuring signals that are rapidly changing, thus the challenges involved with the current manuscript are of a different nature, and require significant testing and integration of a new type of readout circuit, namely the lock-in-amplifier. In the current manuscript, we are integrating two sensors that are of a different class and entirely different nature, a physiological signal sensor which is more of a physical measurement, and a microfluidic impedance sensor, which falls under the class of biological/chemical measurements. In this manuscript, we, for the first time ever, show the potential for using Analog Signal compression for signals obtained from these two different classes of signals. For this purpose, we performed novel experiments to achieve simultaneous microfluidic impedance cytometry and physiological signal sensing. We also performed systematic wireless link level simulations to evaluate the signal recovery performance of the proposed wireless biosensing system under various indoor and outdoor signal propagation conditions.  This work shows the promise of our vision for a compression scheme that would work with a network of body sensors that are continuously obtaining both physical and biochemical data.

\textbf{Our Contributions:}
In this study, we mimic the measurement of blood cells on a microfluidic impedance cytometer (microfluidic channel with \SI{30}{\micro\meter} width and \SI{20}{\micro\meter} height) using \SI{7.8}{\micro\meter} diameter beads in saline buffer. The physiological signals are the electrical Galvanic Skin Response~(GSR) measurement of skin conductance via Shimmer sensors with a maximum value of \SI{2.6}{\Omega^{-1}}. This work is distinct from prior works in that the detected physiological and biomarker analog signals are directly compressed in the analog domain by an analog circuit, and are modulated and transmitted via analog Radio Frequency~(RF) communication chain. There are no power-hungry digital circuits such as ADCs, Digital-to-Analog Converters~(DACs) or microprocessors in the sensor, which enables the sensor to run on power gathered from energy-harvesting techniques such as human motion. The all-analog circuit performing the compression is based on an encoding technique termed Analog Joint Source Channel Coding~(AJSCC)~\cite{Zhao16, wons3tier2017,MASS17}, where a digital Cluster Head~(CH) receiver performs the AJSCC decoding to recover the source signals. 
The major contributions of this work include: 
\begin{itemize}
\item An all-analog wireless biosensor design is presented, consisting of concurrent microfluidic impedance sensing as well as physiological signal sensing and subsequent signal compression by an AJSCC circuit.
\item Two circuit designs of the AJSCC encoding are presented---the first supporting a fixed and small number of AJSCC levels, and the second supporting a flexible and larger number of levels (within a certain range).
\item The presented biosensor system is validated via hardware experiments using the first design (Design~1), and by circuit and wireless link simulations for the second design (Design~2).
\end{itemize}

\textbf{Article Outline:}
In Sect.~\ref{sec:sensor}, our wireless biosensor design for dual measurements is presented including its equivalent circuit model, microfluidic device fabrication procedure, and improved design for AJSCC encoding circuitry. In Sect.~\ref{sec:perfeval}, hardware experimental results for Design 1, circuit simulation and wireless link simulation results for Design 2 are presented. Finally, in Sect.~\ref{sec:conc}, conclusions are drawn.

\begin{figure*}
\begin{center}
\includegraphics[width=7.0in]{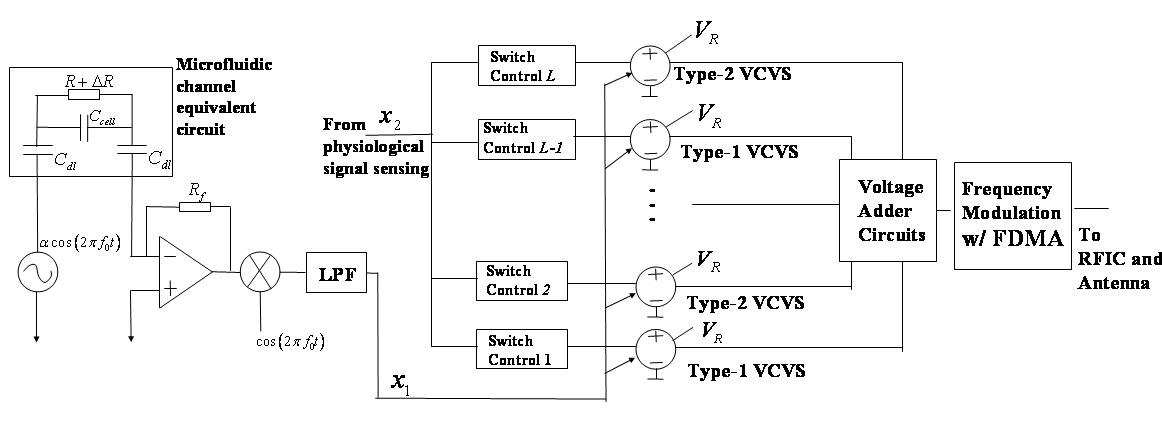}
\end{center}
\caption{All-analog circuit diagram with microfluidic and physiological sensing signal compressed by Analog Joint Source Channel Coding~(AJSCC) circuit~\cite{Zhao16} with Frequency Division Multiple Access~(FDMA) to multiplex a large number of sensors. The microfluidic equivalent circuit and the lock-in amplifier front-end circuit are depicted. The AJSCC circuit realizes fixed number of levels with stacked VCVS design (Design~1). Note that, the time-domain waveforms of microfluidic signal $x_1$, physiological signal $x_2$ and the AJSCC-encoded signal are depicted in Fig.~\ref{fig_output_Tx}.}\label{fig_Proposed_Biosenor}
\end{figure*}

\section{Proposed Wireless Biosensor Prototype}\label{sec:sensor}
We first present the system design of the wireless biosensor prototype including the equivalent circuit model of the microfluidic channel, the prototype system setup, and the microfluidic device fabrication procedure in Sect.~\ref{sec:sensor:system}. Then, an improved circuit design of the AJSCC encoding, Design~2, is presented in Sect.~\ref{sec:sensor:design2}.

\subsection{Wireless Sensor System}\label{sec:sensor:system}

The wireless sensor circuit depicted in Fig.~\ref{fig_Proposed_Biosenor}, is composed of the microfluidic sensing circuit, and the all-analog signal compression and wireless transmission circuits. The microfluidic system adopts impedance-based particle-concentration measurement. The equivalent circuit of the impedance-based microfluidic system in this study has been used for prior work also in~\cite{Emaminejad12}. The microfluidic device consists of the microfluidic PDMS (polydimethylsiloxane) channel on a pair of co-planar electrodes. The equivalent circuit inside the microfluidic channel and the sensor can be modeled by two double-layer capacitors $C_{dl}$, one resistor $R$, and one capacitor $C_{cell}$ in parallel as shown in Fig.~\ref{fig_Proposed_Biosenor}. When a molecule passes by, the resistor will have a change in resistance $\Delta R$, which will result in a change in signal detection. The impedance measurement is excited with a cosine wave with frequency $f_0$. The resistance change $\Delta R$ and the voltage change will be detected and amplified by the lock-in amplifier with feedback resistance $R_f$. The signal is then passed to a mixer with frequency $f_0$ and then through a low-pass filter. For this equivalent circuit model, molecules passing the electrode will cause impedance variation, and molecules of different sizes will have different impedance variation in the output pulse response. The sensitivity depends on a number of conditions, including the dimensions of the microfluidic channel, the choice of the excitation frequency, the noise level of the front-end circuit, and the detection algorithm. An optimized impedance-based microfluidic system considering all of these parameters can detect sub-$\mu\mathrm{m}$ particles. In our previous work, we have shown that the microfluidic sensor can detect micron-sized beads variations within \SI{2}{\micro\meter} in diameter (adequate for cells). In other studies, researchers have demonstrated the ability to detect particles at the nano-scale~\cite{balakrishnan2013node,le2016biomems}.

\textbf{System Setup:}
To demonstrate a proof-of-concept design, we quantify the microbeads in the microfluidic channel to mimic the blood cell quantification process. In this work, the sensor quantifies the number of beads, which is similar to how the number of cells in a blood sample would be quantified. The detection of real cells in a complex biological sample like blood is beyond the scope of this work.
The cells are loaded into the micro-channel and directed to the microfluidic sensing region. The particles passing the electrode will produce pulses in the voltage output as described in~\cite{Lin15,Emaminejad12}. The microfluidic signal is generated by micron-sized beads passing through the co-planar electrode pair in the microfluidic channel with a width of \SI{30}{\micro\meter} and a height of \SI{20}{\micro\meter}. The electrode dimension is \SI{20}{\micro\meter} width and \SI{200}{\nano\meter} length, with a spacing of \SI{25}{\micro\meter} between the electrode pair. The effective length of the electrodes is \SI{30}{\micro\meter}, which is the amount overlapping with the microfluidic channel and is exposed to the electrolyte. The signal is excited with cosine-ware frequency $f_0$ of \SI{500}{\kilo\hertz}. The flow rate is \SI{0.1}{\micro\liter/\minute}. The bead size is \SI{7.8}{\micro\meter}. A sequence of pulses is generated with beads flowing over the electrode pair and the signal is recovered using the lock-in amplifier analog circuit~\cite{Talukder17}.

\textbf{Microfluidic Sensor Fabrication:}
The microfluidic impedance cytometer is fabricated using soft lithography~\cite{xia1998soft}. The sensor and the microchannel are fabricated individually before incorporating into the microfluidic device. Photomasks for the lithography processes are designed using AutoCAD. The photomasks are fabricated by Advance Reproductions Corp. (North Andover, MA).\\
\textit{(i)~Microfluidic Channel:}
The microchannel is constructed using PDMS and standard molding techniques~\cite{xia1998soft}. The master mold is fabricated with SU-8 photoresist. To fabricate the microchannel, PDMS solution is poured onto the mold and cured at $\SI{80}{\celsius}$ for an hour. After curing, the PDMS is peeled off from the mold to create the microfluidic channel. \\
\textit{(ii)~Microelectrode:}
The microelectrodes are fabricated on the glass substrate using standard photolithography. Photoresist AZ5214 is spin coated on the substrate and exposed to UV light under the the microelectrode photomask. The coated glass wafer is then developed in AZ5214 developer. The photoresist exposed to UV light would be washed away creating the pattern of the microelectrodes. A thin layer of chromium and gold ($\SI{50}{\angstrom}$ and $\SI{200}{\nano\meter}$ respectively) are deposited on the wafer using using electron beam evaporation. \\
\textit{(iii)~Microfluidic Device:}
A PDMS chip embedded with a microchannel is covalently bonded to the glass wafer with microelectrodes to create the microfluidic device. The surfaces of glass and PDMS are exposed to oxygen plasma to generate thin layers of silanol terminations~(SiOH). When brought in contact with the oxidized glass surface, the silanol terminated layers come together to create the conformal Si-O-Si covalent bonds between the polymer and glass~\cite{duffy1998rapid}. The oxygen plasma also makes the PDMS surface hydrophilic~\cite{tan2010oxygen}. This allows for capillary flow of fluid/cells through the impedance cytometer without the need for external syringe pumps.

\textbf{Key Sensor Circuit Design~1}:
The circuit design (Design~1) of the AJSCC encoding is depicted in Fig.~\ref{fig_Proposed_Biosenor}. In the all-analog realization of AJSCC circuit~\cite{Zhao16}, the microfluidic sensing signal $x_1$ controls the output of Voltage Controlled Voltage Sources~(VCVS). Note that there are two types of VCVS, type-1 and type-2. For type-1 VCVS, the output voltage of VCVS increases linearly with the increment of the controlling voltage $x_1$, which corresponds to the odd-numbered levels in AJSCC; and for type-2 VCVS, the output voltage of VCVS decreases linearly with the increment of the controlling voltage $x_1$, which corresponds to the even-numbered levels in AJSCC. Each VCVS is switched among saturation voltage $V_R$, the linear voltage output corresponding to $x_1$, and ground. The physiological signal $x_2$ is the control signal of the VCVS composed of $L$ stages. With larger physiological signal, there will be more stages being activated. For example, if there are $M$ stages being activated, the $M$-th stage will be controlled by the signal coming from the impedance cytometer $x_1$ to produce a continuous varying output voltage, and the $1$-st to the $M-1$-th stages will output the maximum voltage $V_R$. The other higher $L-M$ stages produce zero grounded outputs. The voltages of all the stages are summed together by an analog voltage adder to produce the desired AJSCC encoded voltage. The signal is then passed to RFIC and the antenna. Because of the stacking of multiple VCVS blocks corresponding to different stages, we call this ``Stacked VCVS'' design.

The power consumption of our analog AJSCC board (with ``Stacked VCVS'' design) without radio power is estimated to be $130~\rm{\mu W}$ for a discrete-component realization as follows~\cite{Zhao16}. Our circuit in total (5 and a half stages/11 levels) consists of 16 OpAmps, 17 Comparators, and 11 Multiplexers, where OpAmps are clearly the major contributors to the overall power consumption. There are many low-power designs proposed for these components. For example, a low-power OpAmp~\cite{opamp} consuming about $8~\mathrm{\mu W}$, a comparator~\cite{comparator} consuming about $12.7~\mathrm{nW}$, and an analog multiplexer ($ADG704$) consuming about $10~\mathrm{nW}$ can be used for our circuit resulting in a power consumption of $\approx\mathrm{130 \mu W}$. We believe this number can be even lower (to less than $50~\rm{\mu W}$) if an Integrated Circuit~(IC) design is adopted, because of the following reasons: (i)~a design using discrete components for different functionalities of the circuit uses extra hardware as inter-component optimization is not feasible as in an IC implementation; (ii)~in general, the larger the area of the chip, the higher the power consumption~\cite{discrete_ic}, which is due to the use of a larger substrate or a combination of several smaller substrates (corresponding to different discrete components). On the other hand, in an IC, all the functionalities are designed on top of a single tiny substrate resulting in much lower area and, hence, a much lower power consumption; (iii)~using IC design allows the use of latest nm-Si (nanometer-Silicon) technology (e.g., $10~\mathrm{nm}$ currently or $5~\mathrm{nm}$ in near future%
) for the fabrication of the chip, which again results in lower power consumption as the area is reduced owing to lower dimensions of the transistors' channel. We expect these reasons will help reduce the power consumption by more than half resulting in $< 50~\rm{\mu W}$ power consumption. On the other hand, for state-of-the-art wireless digital sensors (e.g.,~\cite{wsn430,telosb}) the consumption is at least $1$-$2~\rm{mW}$ in active mode. A detailed power comparison of our AJSCC system with existing wireless sensors can be found in~\cite{wons3tier2017}. The fabrication cost can also be kept lower than the digital counterpart by leveraging mass IC production.

\subsection{Key Sensor Circuit Design~2}\label{sec:sensor:design2}
There is a scope of further reduction in power consumption and circuit complexity (which translates to cost) of the AJSCC circuit in Design~1 (Sect.~\ref{sec:sensor:system}). As can be noticed, Design~1 adopts a fixed number of AJSCC levels which needs to be decided before the fabrication of the chip, making the sensing resolution in $x_2$ fixed once the chip is fabricated. On the other hand, it may be necessary to vary the sensing resolution based on the application. For example, until certain symptoms are detected a lower resolution in $x_2$ (i.e., higher $\Delta$) can be adopted and the resolution can be increased if some preliminary symptoms are detected. Secondly, in Design~1, the hardware corresponding to each level/stage is duplicated as many times as the number of AJSCC levels. This is an inefficient approach, both in terms of power consumption as well as complexity/cost as the hardware in each level can be reused instead of being duplicated. Finally, since having a large number of AJSCC levels is tedious with this approach (since it does not scale well with the number of levels), it leads to more quantization in $x_2$ (as can be observed in the experimental section, Sect.~\ref{sec:perfeval:exp}).

In order to address the above limitations of Design~1, we present a novel Design~2 where the number of AJSCC levels can be user defined within a certain range (even after the fabrication of the chip) as well as has much lower power consumption and circuit complexity compared to Design~1 (especially for number of levels greater than 16). The improved sensor circuit design (Design~2) is given in our previous work ~\cite{SensorsJournal18}. To connect this circuit with biosensing, the physiological signal $x_2$ is first divided by the tunable spacing between levels, $\Delta$ using a multi-stage analog divider to produce the quotient of the division. The quotient is fed to two blocks---an analog multiplier and an odd/even detector block which detects whether the quotient is odd or even. On the other hand, the microfluidic signal, $x_1$ is fed to both type-1 VCVS and type-2 VCVS blocks to produce respective type-1 or type-2 outputs which are proportional and inversely proportional to $x_1$ respectively. Whether to use type-1 output or type-2 output depends on which \textit{numbered} level the mapped AJSCC point is (where the \textit{number} is the quotient of the division). If it falls on an even level ($0, 2,...$), type-1 output needs to be considered and vice-versa. Hence, the odd/even detector (of quotient) can control which output is considered via an analog switch. Then the quotient multiplied by the level saturation voltage, $V_R$ is added to the switch output to produce the final AJSCC encoded voltage.

We have implemented the entire encoding circuit of Design~2, i.e., from inputs $x_1$ and $x_2$ till the output of adder using Spice~\cite{SensorsJournal18}, where, the multi-stage analog divider and odd/even detector block combined are implemented as a single circuit. For further details on the implementation, comparison with Design~1 in terms of complexity and power consumption, maximum and minimum variation in $\Delta$ (in other words, the number of levels, $L$) supported by the circuit, please refer to our previous work~\cite{SensorsJournal18}. The choice of the parameter $L$, depends on the distribution of the source signals, the transceiver design, and the wireless channel conditions. To find the optimal parameter $L$, we need to evaluate the wireless communication system performance under different channel conditions, which has been done in Sect.~\ref{sec:perfeval:sim}. From these results, we are able to make the judgment on the optimal $L$ for a given biosensing application.

\begin{figure*}[ht]
        \centering
        \hspace{-0.4in}
            \begin{subfigure}[b]{0.32\textwidth}
         		\centering
        		\includegraphics[width=1.15\textwidth]{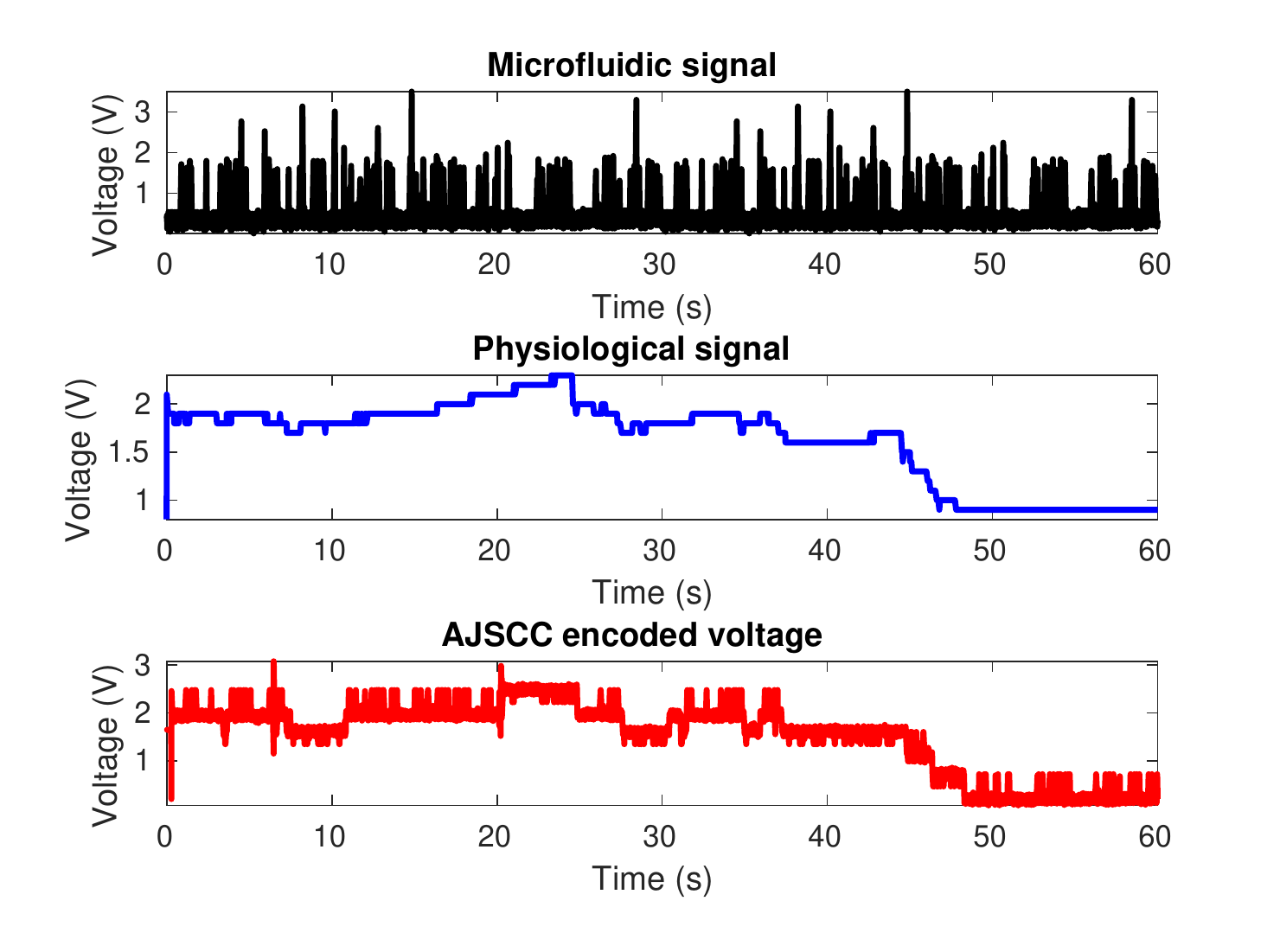}
         		\caption{}
         		\label{fig_output_Tx}
         	\end{subfigure}
~
        \begin{subfigure}[b]{0.32\textwidth}
            \centering
            \includegraphics[width=1.15\textwidth]{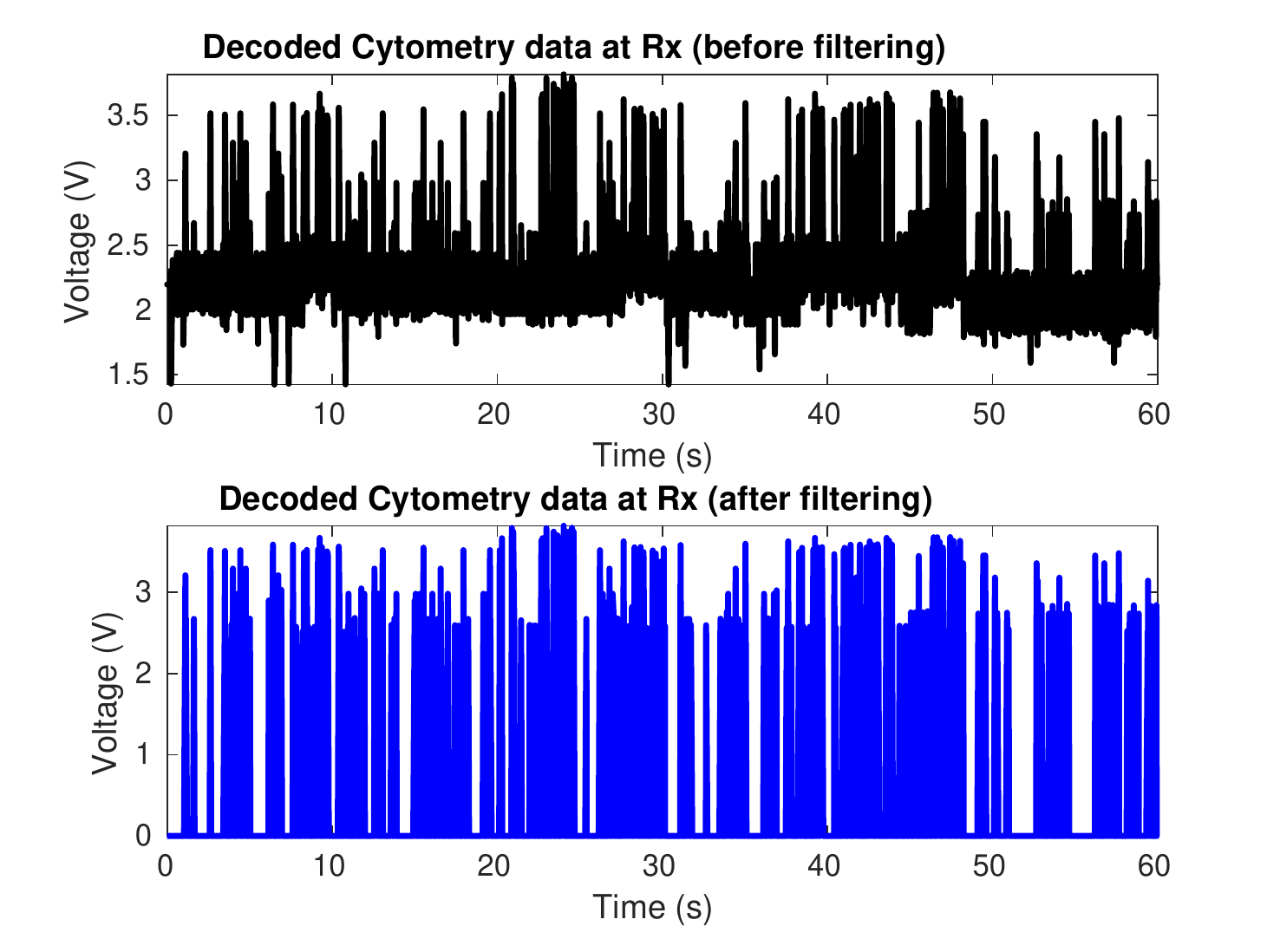}
            \caption{}
            \label{dec_cytometry}
        \end{subfigure}
~
         \begin{subfigure}[b]{0.32\textwidth}
             \centering
             \includegraphics[width=1.15\textwidth]{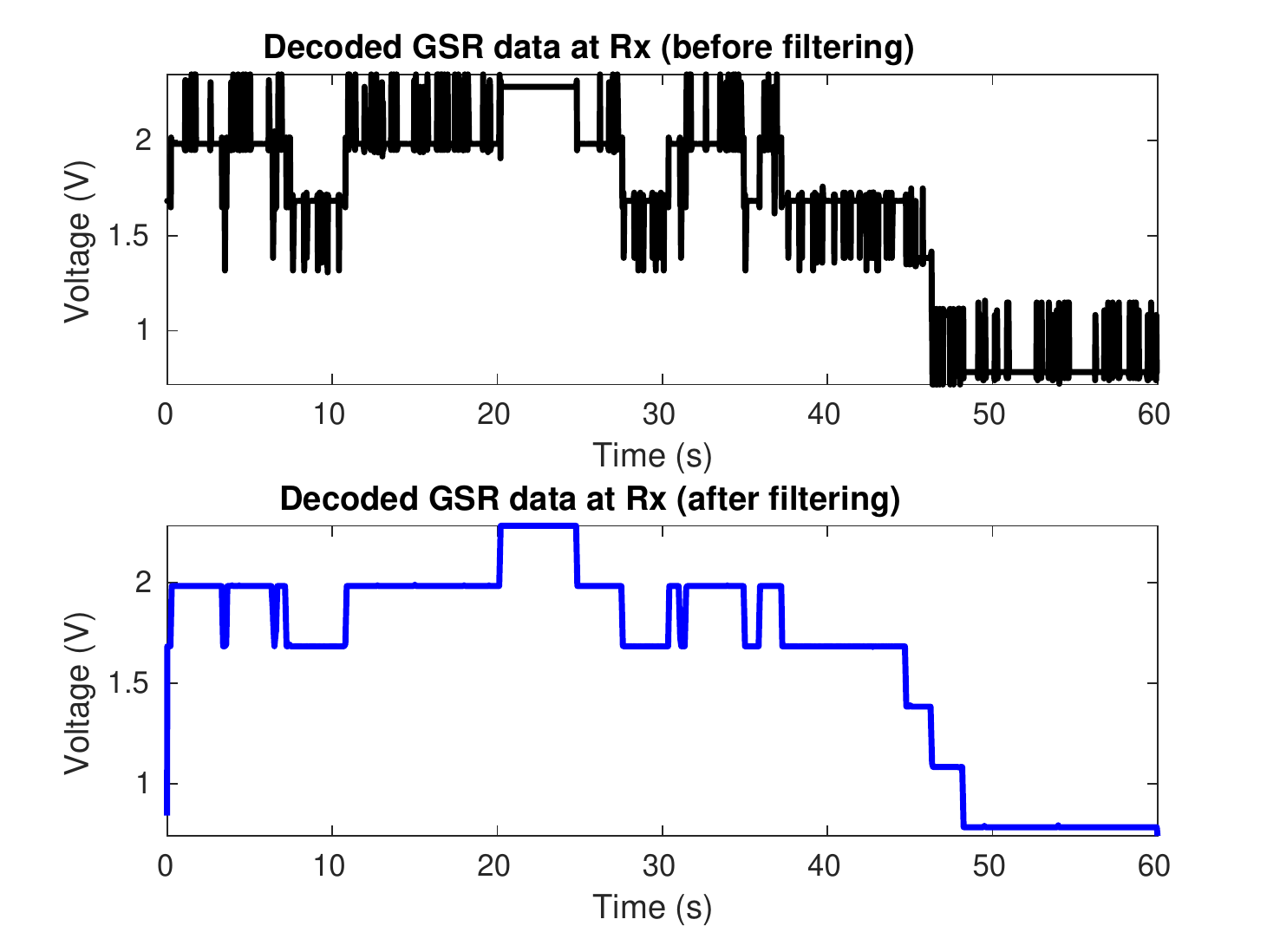}
             \caption{}
             \label{fig_dec_gsr}
         \end{subfigure}
         \caption{(a)~Microfluidic (cytometry), physiological (GSR), and AJSCC-encoded voltage signal at the transmitter. With respect to Fig.~\ref{fig_Proposed_Biosenor}, $x_1$ is the microfluidic (cytometry) signal, $x_2$ is the physiological (GSR) and the output of the AJSCC encoding circuits is the AJSCC-encoded voltage. (b)~Decoded microfluidic (cytometry) signal at the digital CH receiver---before~(top) and after~(bottom) filtering; (c)~Decoded physiological~(GSR) signal at the CH---before~(top) and after~(bottom) filtering. The bead site is \SI{7.8}{\micro\meter} and the flow rate is \SI{0.1}{\micro\liter/\minute}. The microfluidic channel has width of \SI{30}{\micro\meter} and height of \SI{20}{\micro\meter}. The GSR signal (measured as skin conductance) is measured with the Shimmer GSR sensor with maximum value of $2.6~\mathrm{M}\Omega^{-1}$.}
\end{figure*}

\begin{figure}
\begin{center}
\includegraphics[width=3.5in]{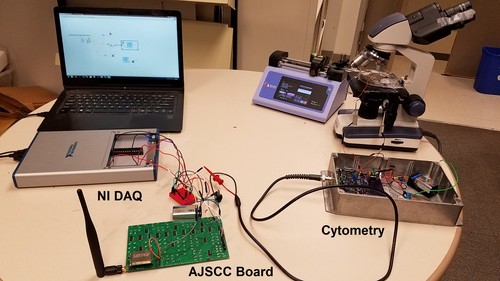}
\end{center}
\caption{The right-hand side shows the cytometry system, which generates the impedance cytometry signal. Prior collected GSR data is used to generate an analog signal using Labview on a computer and NI DAQ system~(left top). Both these analog signals are fed to the AJSCC board~(left bottom), which does the AJSCC encoding and transmits wirelessly to a digital CH receiver (not shown). The impedance pulses are generated by the impedance cytometer, and the GSR signals are generated by the NI DAQ device. The AJSCC board performs data compression in the analog domain, and transmits the signal wirelessly.}\label{fig_system}
\end{figure}

\section{Performance Evaluation}\label{sec:perfeval}
In this section, we present the validation results of the proposed biosensing platform. Specifically, we present the experimental results of our wireless biosensor using the AJSCC Printed Circuit Board~(PCB) implementation of Design~1 in Sect.~\ref{sec:perfeval:exp}. 
Then, the circuit simulation results of Design~2 and wireless link simulation results of the sensor with Design~2 to find the optimal parameters are presented in Sect.~\ref{sec:perfeval:sim}.

\subsection{Experimental Results with Design~1}\label{sec:perfeval:exp}
The experiment is designed as a proof of concept to verify the functionalities of the key sensor circuit Design~1 (11 AJSCC levels using stacked VCVS design) at the transmitter side as well as at the receiver to recover the two compressed signals.

\textbf{Experimental Setup:}
The test is performed in an indoor environment with a carrier frequency of $2.4~\rm{GHz}$ and an omni-directional antenna. The distance between the transmitter and receiver is about 5 meters. The transmitter system is shown in Fig.~\ref{fig_system}. The physiological signal is chosen to be a GSR signal connected via Bluetooth using Shimmer sensors, and the signal is regenerated using NI LabView/DAQ system. Both of these signals are fed to the circuit board developed in~\cite{Zhao16} to perform the sensor signal encoding in the analog domain.

\textbf{Transmitter AJSCC Encoding:}
The original microfluidic signal, physiological signal, and the AJSCC encoded signal in the transmitter are depicted in Fig.~\ref{fig_output_Tx} as captured using NI DAQ hardware and LabView software. 

We can observe that AJSCC encoding compresses the two signals into a single one such that the microfluidic signal is riding on top of the physiological signal. We note that our AJSCC sensor board introduces a quantization error in the physiological signal since it has only 11 levels in the y-dimension, as seen in Fig.~\ref{fig_output_Tx} where the physiological signal portion in the AJSCC encoded signal has been quantized. This AJSCC encoded signal is frequency modulated, upconverted, and then transmitted using Commercial Off The Shelf~(COTS) RFIC chip. %

\textbf{Receiver Decoding:}
At the receiver, the signal is downconverted (to baseband) and then frequency demodulated to recover the AJSCC encoded signal. The demodulation process is as follows. The baseband signal is captured into LabView using an NI DAQ device. Then the signal is firstly sampled with a sampling rate $f_s$, then the frequency of the signal is detected by an Fast Fourier Transform~(FFT)-based frequency detector with $N_s$ samples per FFT. At the receiver, the parameter $N_s$ is chosen as $5,000$ and sampling rate $f_s$ is set to $500~\rm{kHz}$, which is the maximum supported by the DAQ device. The AJSCC-encoded signal is recovered from the frequency values using simple linear mapping (as done in the transmitter board). The AJSCC-encoded voltage is then decoded to individual physiological and microfluidic signals using simple modulo arithmetic~\cite{Zhao16}. The decoded microfluidic signal in LabView is shown in Fig.~\ref{dec_cytometry}~(top). It can be observed that the peaks of the signal are recovered. The error at the bottom of the signal is due to the bias in the AJSCC stage mapping. The decoded physiological signal is shown in Fig.~\ref{fig_dec_gsr}~(top). The quantization effect of the recovered physiological signal is due to the fact that in the transmitter board only 11 stages are designed for the AJSCC mapping. The error floor at the bottom of the microfluidic signal is removed using a thresholding filter, where the threshold value is set just above the error floor level. The unwanted spikes in the physiological signal are smoothed using a $200$-th order median filter. The bottom portions of Figs.~\ref{dec_cytometry} and \ref{fig_dec_gsr} show the results after filtering.

Scaling of physiological and impedance cytometer values is done as follows: physiological being the conductance measurements have the range in $100$'s of $k\Omega^{-1}$ below 2.6$M\Omega^{-1}$, which is not suitable value to be fed to hardware, hence we linearly scaled the physiological values to a range from $1$ to $3~\rm{V}$. On the other hand, even though microfluidic values are within acceptable voltage ranges ($0-2~\rm{V}$), their contribution reduces significantly in the AJSCC encoded signal due to $1:5$ Voltage Controlled Voltage Source~(VCVS) mapping in the AJSCC encoding process~\cite{Zhao16}. Hence we had to upscale the microfluidic signal to $0-10~\rm{V}$ to make the microfluidic signal and its peaks apparent in the AJSCC encoded signal. 
In the experiments, the NI~DAQ device was sampling the received baseband signal at $500~\rm{kHz}$; however, the NI LabView was not able to process the data at the same rate. Hence, we collected the samples alone in the LabView (without further processing) and then processed the samples in MATLAB for frequency detection and AJSCC decoding. This also gave us the opportunity to vary the window size $N_s$ (and find the best value) for frequency detection: a small window means fewer samples for frequency estimation resulting in wrong frequency values; on the other hand, a large window results in loss of sampling accuracy (as we get averaged values instead of actual values). %
After trying different values, we found that $N_s$=$5,000$ gives the best results in that it matches closely the original physiological and microfluidic signals.

\begin{figure}
\begin{center}
\includegraphics[width=3.5in]{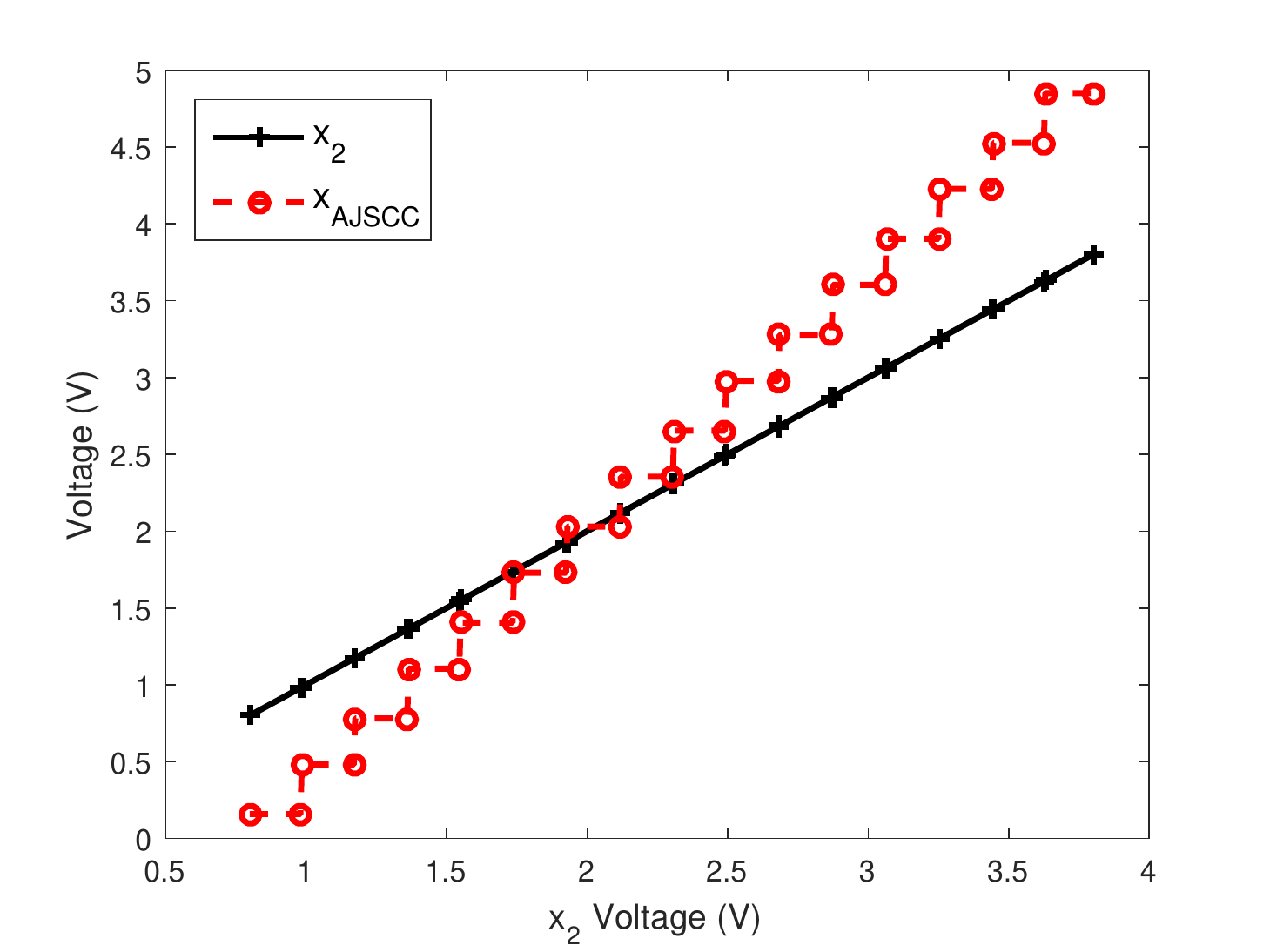}
\end{center}
\caption{AJSCC-encoded output of 16-level analog divider circuit with varying $x_2$ and $x_1 = 2.5~\mathrm{V}$ and $V_{REF}=3~\rm{V}$ by LTSpice circuit simulation.}\label{fig:LevelsPlot}
\end{figure}

\subsection{Simulations of Design~2}\label{sec:perfeval:sim}
In this subsection, we first briefly present the simulation results of the improved AJSCC encoding circuit (Design~2, Sect.~\ref{sec:sensor:design2}). Then the proposed wireless biosensor system Design~2 that allows a large number of AJSCC levels is evaluated in terms of the wireless channel conditions via wireless link simulations. 

\textbf{Improved AJSCC Circuit (Design~2):}
Fig.~\ref{fig:LevelsPlot} shows LTSpice simulation output of the improved AJSCC circuit's output voltage (AJSCC encoded voltage) while varying $x_2$ and fixing $x_1$ at $2.5~\rm{V}$ so that the mapped point will always be at the center of each level. The figure shows all the 16 levels possible with $k=4$ and matches the expectations of the circuit in that---we can notice a discretization in the AJSCC encoded voltage, $x_{AJSCC}$ owing to the discretization of $x_2$ by the AJSCC encoding circuit. However, the amount of discretization depends on the number of AJSCC levels and Design~2 enables the realization of a large number of levels easily as compared to Design~1 (the result shown here is specifically for 16 levels). We describe the setup used for wireless link simulations of our wireless biosensor system with the flexible number of AJSCC levels (i.e., adopting Design~2).

\textbf{Wireless Link Simulation Setup:}
The same cytometry and GSR data used in the hardware experiments (Sect.~\ref{sec:perfeval:exp}) is adopted in the wireless link simulator for evaluating the system performance. The cytometry and GSR data both have sample spacing of $1~\rm{ms}$. The AJSCC encoder in the simulator has the same voltage range for $x_1$ and $x_2$ as the hardware PCB board, i.e., the $x_1$ has a maximum of $2.25~\rm{V}$, and $x_2$ has a maximum of $3.0~\rm{V}$. The first step is to adjust the range of input signals, $x_1, x_2$ to the AJSCC encoder. This is done with a linear scaling of the input signals to the input range of AJSCC encoder. We have also implemented the frequency modulation adopted in the hardware board in the transmitter of wireless link simulator. In the receiver, the sampled data is sent to the FFT block for peak detection and signal recovery. Two sets of sampling rate and FFT sizes are adopted in the simulation. The first set is the same set of parameters used in the hardware experiment--\SI{500}{\kilo\hertz} sampling rate and FFT size of $5000$. However, this set of parameters will need a time duration of 10ms to collect the data for producing one sample of the source signal.
The $500~\rm{kHz}$ sampling rate in the hardware experiment is due to the limitation of sampling rate in the NI DAQ device adopted in the experiment. In the simulation, we are able to evaluate at a much higher sampling rate for real-time sensing and processing of the source signal. The second set of parameters is motivated from this objective. This set of parameters is designed to improve the previous time duration to $1~\rm{ms}$. The sampling rate is $8.192~\rm{MHz}$, and the FFT size is 8192. The frequency-modulated signal is mapped to frequency up to $4~\rm{MHz}$. In this simulation, the receiver is able to decode the transmitted signal in real time corresponding to the software-based design of an improved digital receiver.

In the receiver, the analog signal is firstly downconverted to baseband frequency and then sampled using an ADC. The digital signal is then passed to the FFT block for generating the frequency-domain response. Peak detection is subsequently performed to find the frequency-modulated signal to recover the AJSCC-encoded signal. The AJSCC decoding is then performed by reverse mapping on the AJSCC curve to recover the two original mapped source signals. On the choice of the sampling rate, we have evaluated two systems with sampling rates of $500~\rm{kHz}$ and $8.192~\rm{MHz}$ to illustrate the difference in the performance. We can observe that the sampling rate of $8.192~\rm{MHz}$ achieves better Mean Square Error~(MSE) performance compared to a sampling rate of $500~\rm{kHz}$, under the same simulation assumptions.

\begin{figure}
\begin{center}
\includegraphics[width=3.5in]{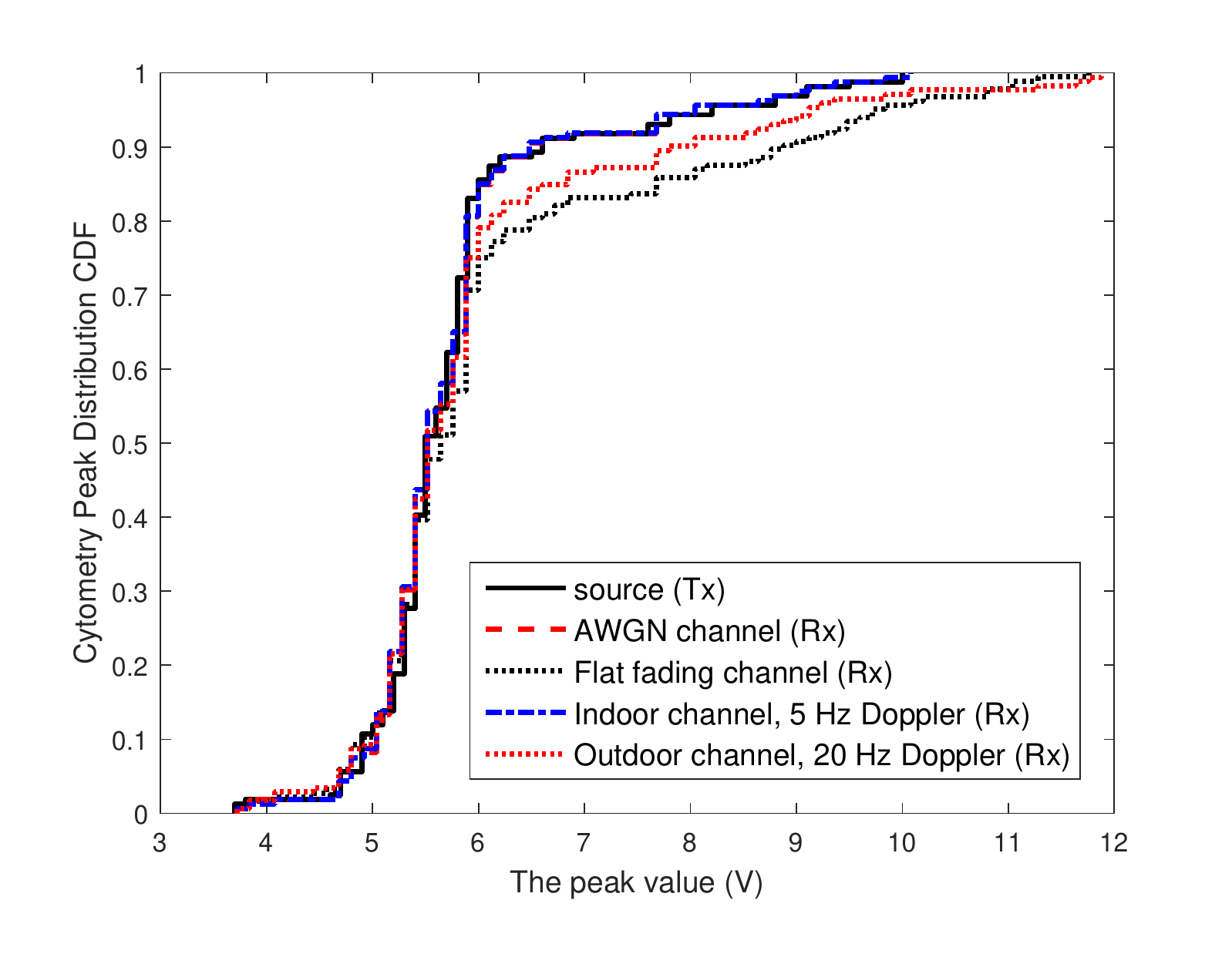}
\end{center}
\vspace{-0.15in}
\caption{Cumulative Distribution Function (CDF) of cytometry peak values at source, at the AJSCC decoding output in receiver for AWGN channel, flat fading channel, indoor channel with 5 Hz Doppler, and outdoor channel with 20 Hz Doppler, by wireless link simulation.}\label{cdf_tx_awgn_flat_indoor_outdoor}
\end{figure}

\begin{figure*}[ht]
        \centering
        \hspace{-0.4in}
            \begin{subfigure}[b]{0.32\textwidth}
         		\centering
        		\includegraphics[width=1.15\textwidth]{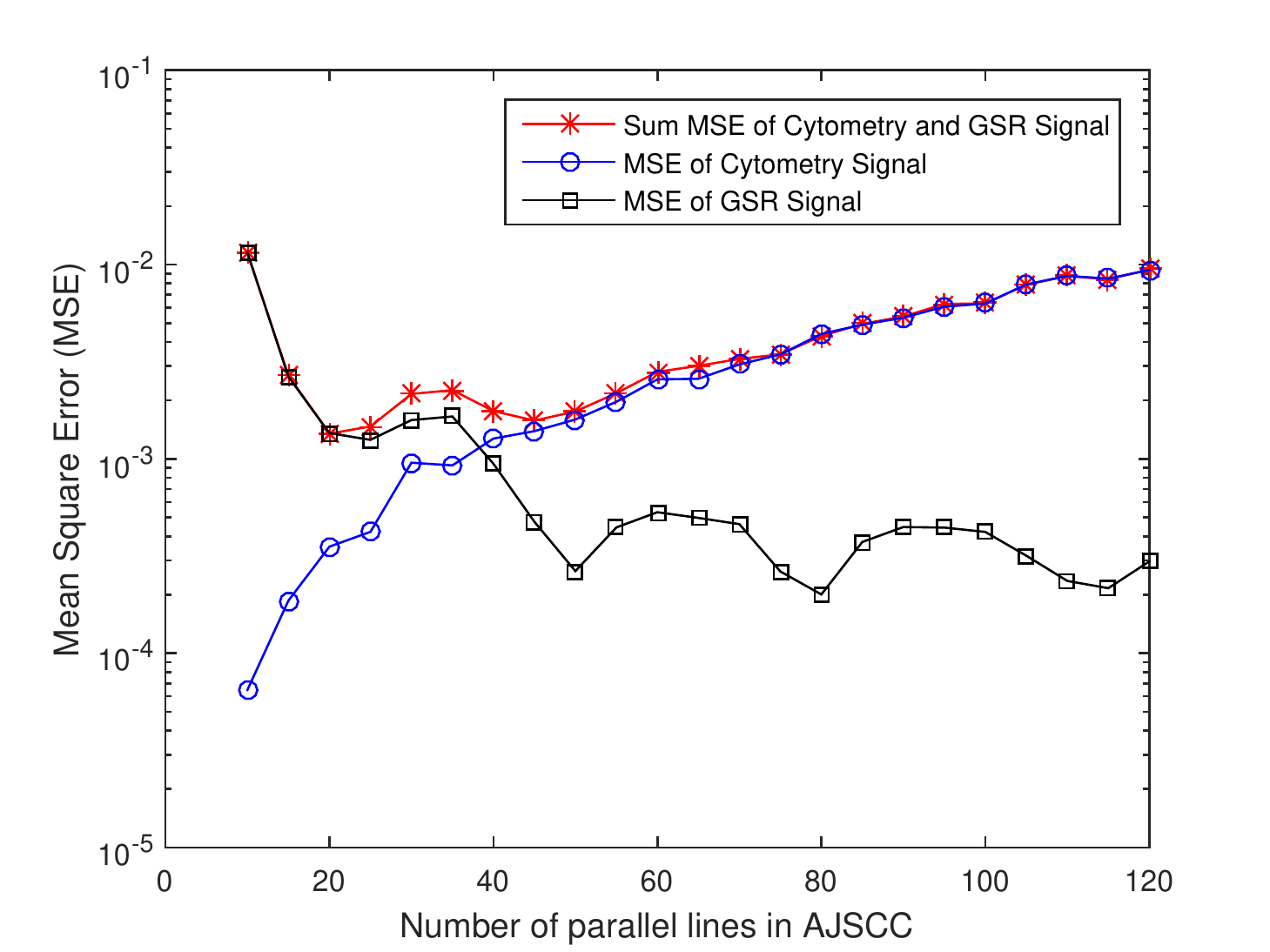}
         		\caption{}
         		\label{fig:MSE_AWGN_left}
         	\end{subfigure}
~
        \begin{subfigure}[b]{0.32\textwidth}
            \centering
            \includegraphics[width=1.15\textwidth]{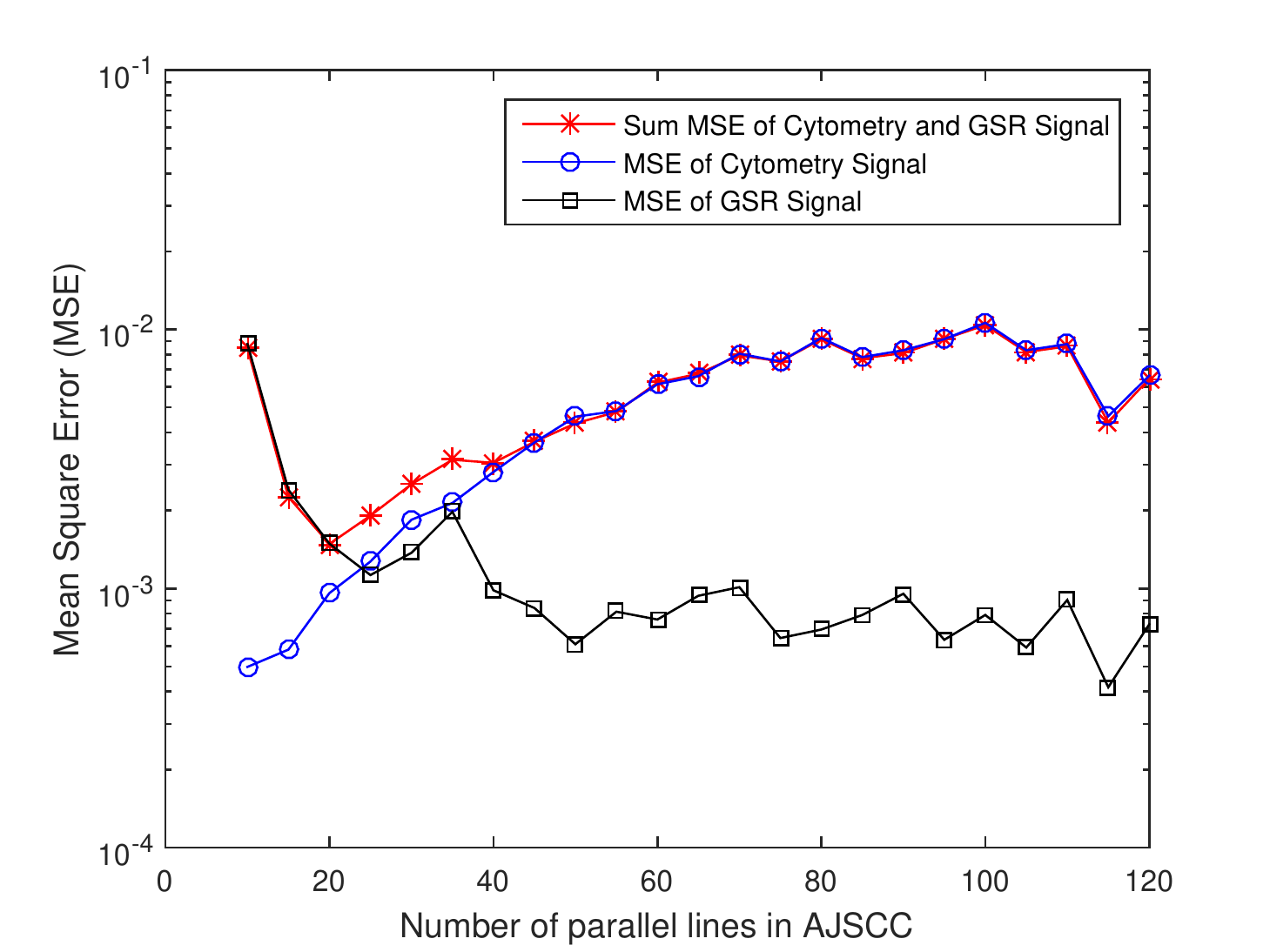}
            \caption{}
            \label{fig:MSE_AWGN_right}
        \end{subfigure}
~
         \begin{subfigure}[b]{0.32\textwidth}
             \centering
             \includegraphics[width=1.15\textwidth]{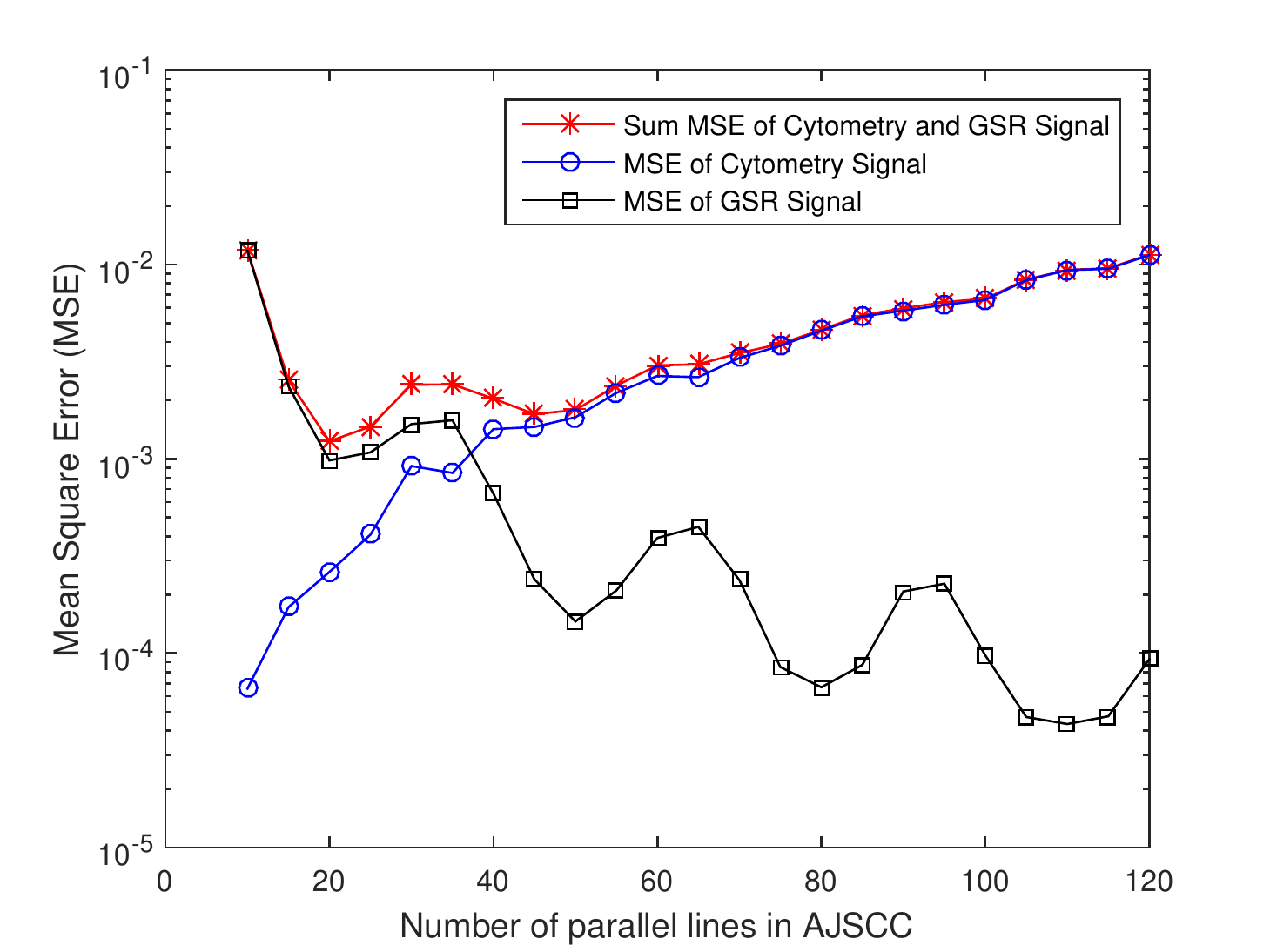}
             \caption{}
             \label{fig:MSE_Ch2}
         \end{subfigure}
         \caption{Mean Square Error~(MSE) vs. number of parallel lines in AJSCC with digital receiver by wireless link simulation, for (a) AWGN channel with CSNR equaling $0~\rm{dB}$, based on digital receiver of ~$8.192~\rm{MHz}$ sampling rate and 8192 FFT size; (b) AWGN channel with CSNR equaling $0~\rm{dB}$, based on digital receiver of  ~$500~\rm{kHz}$ sampling rate and 5000 FFT size; (c) Flat fading channel with zero Doppler and CSNR equaling $0~\rm{dB}$, based on digital receiver of ~$8.192~\rm{MHz}$ sampling rate and 8192 FFT size.}\label{fig:mse_awgn_flat}
\end{figure*}

The channel modeling is critical in the evaluation of wireless system link-level performance~\cite{Zhao07,Sarnoff07}. Four types of wireless channels are realized in the simulator: i)~Additive White Gaussian Noise~(AWGN) channel; ii)~Flat fading channel with single-tap and Rayleigh fading, for modeling isotropic scattering around the receiver. There is no Doppler in these channel models; iii)~Indoor multipath fading channel with Doppler. The channel is named \textit{JTC Indoor Residential A}. JTC, which stands for Joint Technical Committee, is the name of the multipath profile channel models~\cite{Ali94}. This channel is almost flat in our system and simulation setup; iv)~Outdoor multipath fading channel with Doppler $20~\rm{Hz}$. The channel model is named \textit{JTC Outdoor Residential Areas--Low antenna A}. The Doppler spread is assumed as $20~\rm{Hz}$. The reason to choose these channel models is that our hardware experimental system is built with $2.4~\rm{GHz}$ radios and works in an environment similar to Wireless Local Area Network~(WLAN) system, for which JTC channel models have been designed. A parameter named Channel Signal-to-Noise Ratio~(CSNR) is defined for evaluating the effect of noise. It is defined as the ratio of channel gain with signal power, to the noise variance in $\rm{dB}$.

\textbf{Cytometry Pulse Peaks at Source and Receiver:}
The cytometry and GSR source signals are inputs to the AJSCC encoder in the simulator. Fig.~\ref{cdf_tx_awgn_flat_indoor_outdoor} plots Cumulative Distribution Function~(CDF) of pulse peak values of the cytometry signal for the following cases--(i)~at source; (ii)~at the receiver after AJSCC decoding for AWGN channel and flat fading channel. The CSNR of the AWGN channel and flat fading channel are all $0~\rm{dB}$; (iii)~at the receiver after AJSCC decoding for indoor channel with 5 Hz Doppler and outdoor channel with $20~\rm{Hz}$ Doppler. The CSNR of the indoor and outdoor channels are all $10~\rm{dB}$ for these distributions. The digital receivers of all channel conditions evaluated adopt $8.192~\rm{MHz}$ sampling frequency with $8192$ FFT size.
To quantify the similarity of the distributions between the source and received peak values, a statistical test is necessary. The received peak data can be evaluated by the Two-sample Kolmogorov-Smirnov~(K-S) test. The K-S test is run on the two sets of data, the source cytometry peak values, and the received cytometry peak values. The digital receivers evaluated have $8.192~\rm{MHz}$ sampling frequency with $8192$ FFT size for all channel conditions. The results of p-values at different AJSCC levels and the K-S test results are summarized in Table~\ref{Table1}. The AWGN and flat-fading channels have $0~\rm{dB}$ CSNR, and the indoor and outdoor channels have $10~\rm{dB}$ CSNR. From the table, we can observe that the p-values for all the channel conditions are not small, which indicates that there is no statistical significance to the difference between the distributions at transmitter and receiver for all the channel conditions evaluated.

\begin{table}[t]
\centering
\caption{P-values of Kolmogorov-Smirnov~(K-S) hypothesis test results for number of AJSCC parallel lines equaling 30 and 50. The null hypothesis of K-S test is that, the peak distributions at source and receiver are from the same distribution. The K-S test is performed on 5\% significance level.}
\label{Table1}\scriptsize
\begin{tabular}{|p{1.0cm}|p{2.5cm}|p{1.1cm}|p{1.4cm}|p{0.7cm}|}
\hline
\textbf{Case \#} & \textbf{Channel Type}  	& \textbf{No. AJSCC Lines} 	& \textbf{P-value} 		& 	\textbf{K-S Test (5\%)}
\\ \hline
\hline
1 & AWGN	           		& 	30	  &    0.3783    	& 0    		\\ \hline
2 & AWGN	           		& 	50	  &   0.3065     	& 0    		\\ \hline
3 & Flat fading, no Doppler &   30     	&  0.2459      	& 0    		\\ \hline 
4 & Flat fading, no Doppler &   50     	&  0.2534     	& 0    		\\ \hline 
5 & Indoor, $5~\rm{Hz}$ Doppler  	& 	30 		& 0.3974  	& 0					\\ \hline
6 & Indoor, $5~\rm{Hz}$ Doppler  	& 	50 		& 0.2448   	& 0					\\ \hline
7 & Outdoor, $20~\rm{Hz}$ Doppler  & 	30 		& 0.2086  	& 0					\\ \hline
8 & Outdoor, $20~\rm{Hz}$ Doppler  & 	50 		& 0.1898  	& 0					\\ 
\hline
\end{tabular}
\end{table}

\begin{figure*}[ht]
        \centering
        \hspace{-0.4in}
            \begin{subfigure}[b]{0.32\textwidth}
         		\centering
        		\includegraphics[width=1.15\textwidth]{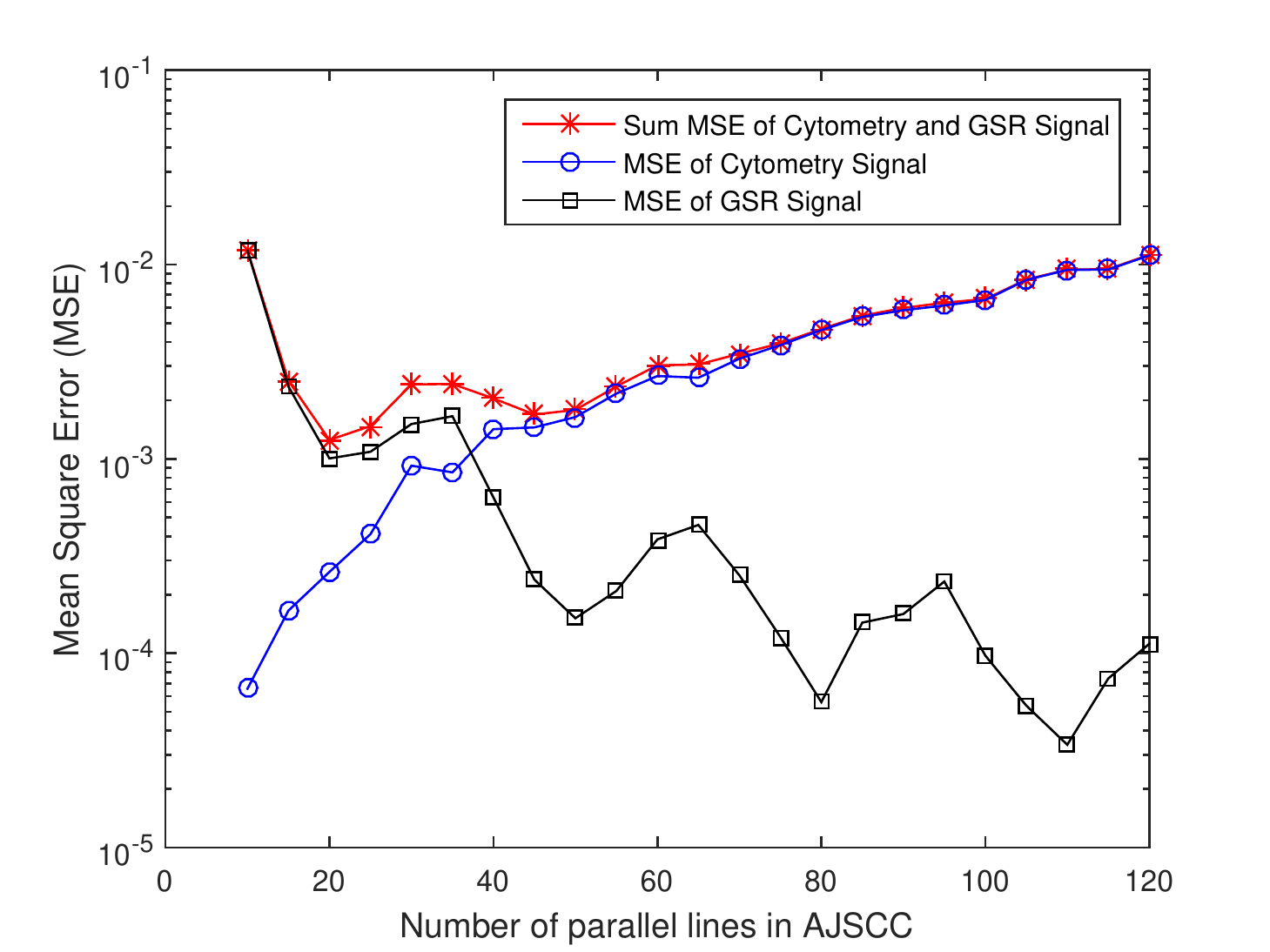}
         		\caption{}
         		\label{fig:CH3_Indoor_FM2_5Hz_0dB}
         	\end{subfigure}
~
        \begin{subfigure}[b]{0.32\textwidth}
            \centering
            \includegraphics[width=1.15\textwidth]{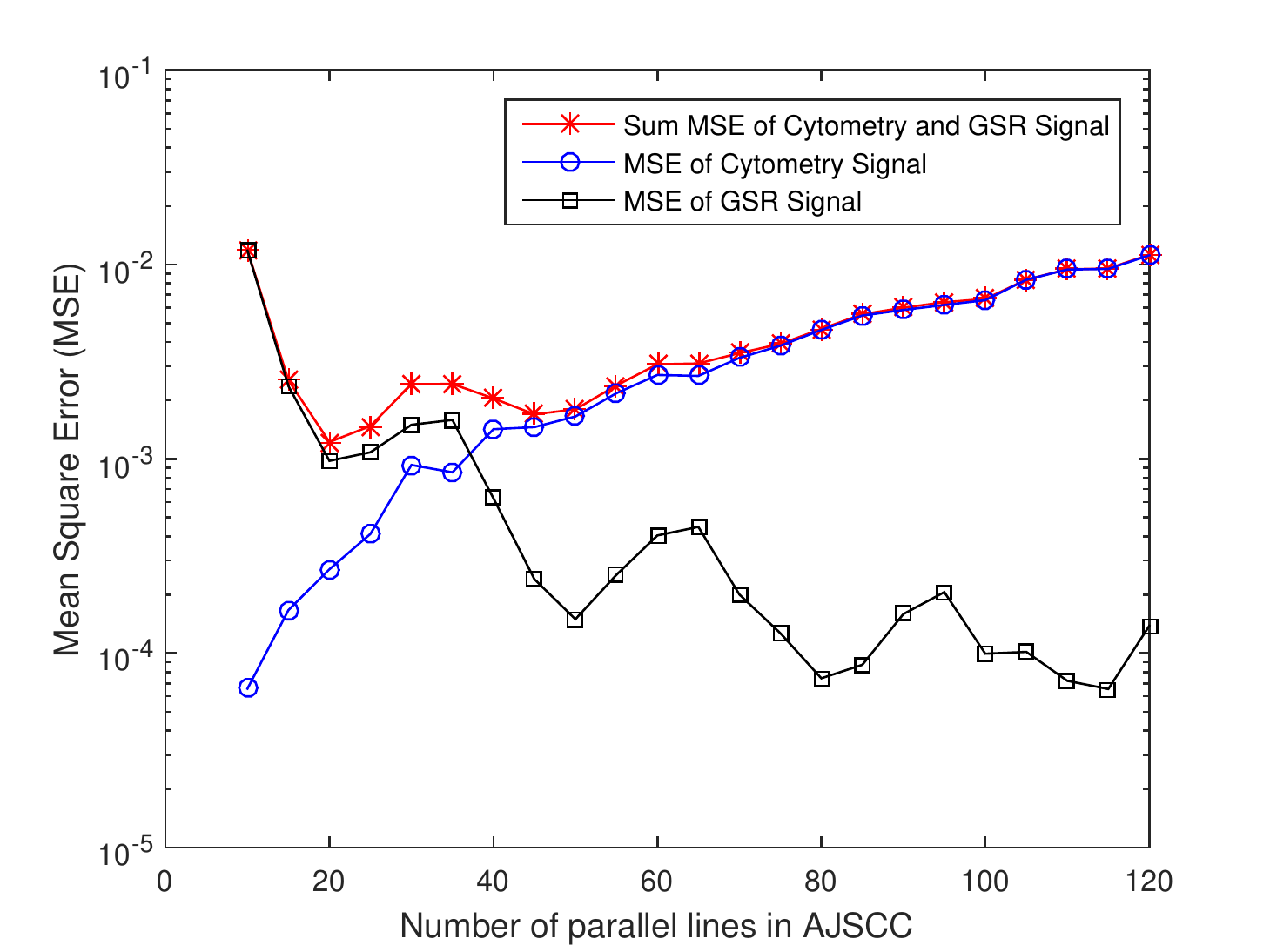}
            \caption{}
            \label{fig:CH3_Outdoor_FM2_20Hz_0dB}
        \end{subfigure}
~		
         \begin{subfigure}[b]{0.32\textwidth}
             \centering
             \includegraphics[width=1.15\textwidth]{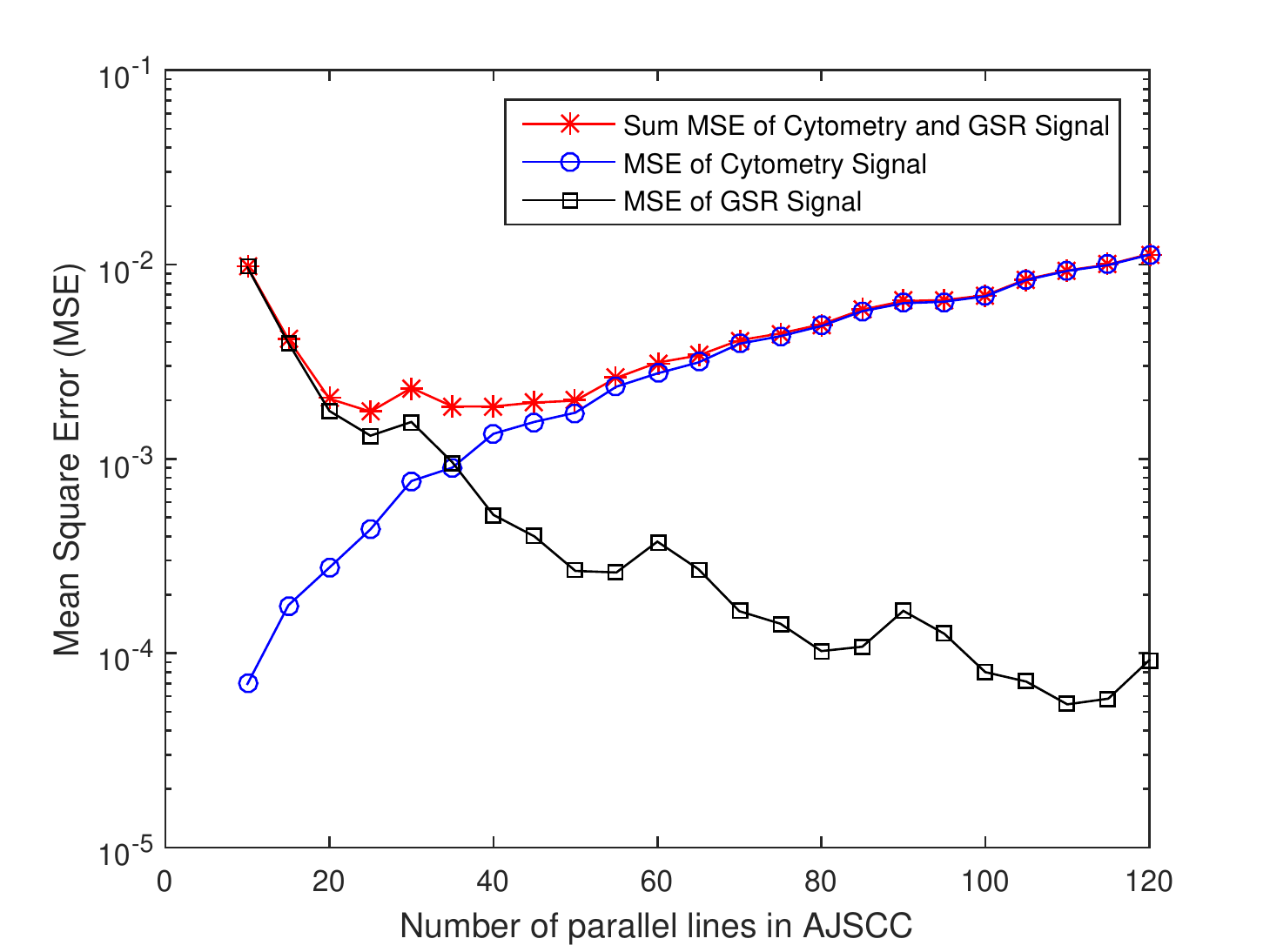}
             \caption{}
             \label{fig:CH3_Outdoor_FM2_20Hz_10dB}
         \end{subfigure}
         \caption{Mean Square Error~(MSE) vs. number of parallel lines in AJSCC for (a)~indoor channel with $5~\rm{Hz}$ Doppler and CSNR equaling $0~\rm{dB}$; (b)~Outdoor channel with $20~\rm{Hz}$ Doppler and CSNR equaling $0~\rm{dB}$; (c)~Outdoor channel with $20~\rm{Hz}$ Doppler and CSNR equaling $10~\rm{dB}$. The digital receivers evaluated are all at ~$8.192~\rm{MHz}$ sampling rate and 8192 FFT size, and results are obtained by wireless link simulation.}\label{fig:mse_indoor_outdoor}
\end{figure*}

\textbf{Signal Recovery Performance of AJSCC:}
MSE performance of the AJSCC is now presented for the proposed system with cytometry and GSR data as source signals. AWGN, flat fading, indoor, and outdoor JTC channels are considered. 

\textit{Performance under AWGN and Flat Fading Channels:}
The simulation results of MSE vs. the number of parallel lines in AWGN channel are depicted in Figs.~\ref{fig:MSE_AWGN_left},~\ref{fig:MSE_AWGN_right}. The channel has CSNR equaling to $0~\rm{dB}$. Two types of digital receivers are evaluated---$8.192~\rm{MHz}$ sampling rate with $8192$ FFT size (Fig.~\ref{fig:MSE_AWGN_left}) and $500~\rm{kHz}$ sampling rate with $5000$ FFT size (Fig.~\ref{fig:MSE_AWGN_right}). It can be observed that the cytometry recovery MSE performance for $8.192~\rm{MHz}$ sampling rate is much lower for a number of AJSCC lines less than or equal to 20 compared with $500~\rm{kHz}$ sampling rate. In addition to real-time processing, this MSE advantage is the second reason that the $8.192~\rm{MHz}$ sampling-rate system is preferred over the $500~\rm{kHz}$ sampling-rate system. The result of MSE vs. the number of parallel lines in AJSCC for flat-fading channel with zero Doppler is depicted in Fig.~\ref{fig:MSE_Ch2}. The results of flat-fading channel are on $8.192~\rm{MHz}$ sampling rate system and CSNR equaling to $0~\rm{dB}$. We observe a trade-off between the MSE of cytometry data and the GSR signal in all plots in Fig.~\ref{fig:mse_awgn_flat}. Since the frequency bandwidth allocated is fixed, with increasing number of AJSCC levels, the frequency resource allocated per level is reduced, therefore the MSE of cytometry data increases with the increase in the number of AJSCC levels; at the same time, with such increase, the accuracy in the quantization of GSR signal is improved and the MSE of the GSR signal reduces. From the curves, we can determine the optimal number of levels in AJSCC that minimizes the sum MSE, which is a large number and hence can be realized via Design~2. We notice that the optimal number of levels depends on channel conditions, source distribution, and transceiver specifications. We can also observe that the sampling rate of $8.192~\rm{MHz}$ achieves better MSE performance compared with the sampling rate of $500~\rm{kHz}$, under the same simulation assumptions. Therefore, in the following simulation evaluations of indoor and outdoor channels, only $8.192~\rm{MHz}$ sampling rate is evaluated.

\textit{Performance under Indoor and Outdoor Fading Channels with Doppler:}
The results of MSE vs. the number of parallel lines in AJSCC for indoor and outdoor fading channels (with same channel models as before) are evaluated. 
All the digital receivers have $8.192~\rm{MHz}$ sampling rate and $8192$ FFT size. The result for indoor channel with $5~\rm{Hz}$ Doppler and $0~\rm{dB}$ CSNR is depicted in Fig.~\ref{fig:CH3_Indoor_FM2_5Hz_0dB}, and the result for outdoor channel with $20~\rm{Hz}$ Doppler and $0~\rm{dB}$ CSNR is depicted in Fig.~\ref{fig:CH3_Outdoor_FM2_20Hz_0dB}. With higher CSNR of $10~\rm{dB}$, the result of outdoor channel with $20~\rm{Hz}$ Doppler is depicted in Fig.~\ref{fig:CH3_Outdoor_FM2_20Hz_10dB}. The result of indoor channel with $10~\rm{dB}$ CSNR is almost the same to the result of $10~\rm{dB}$ CSNR of outdoor channel, hence it is not depicted. From these results, we can observe similar trade-off in MSE of the two source signals for different channel conditions and receiver configurations. These results indicate the optimal number of AJSCC levels for each of the channel conditions simulated. We note that such number varies with channel conditions and transceiver parameters.

\balance

\section{Conclusions}\label{sec:conc}

We presented a new type of wireless biosensing platform to measure molecular biomarkers and physiological signals concurrently, and demonstrated the feasibility of constructing this wireless biosensor. We discussed and compared two designs for the AJSCC encoding circuits. The second design provides a much higher flexibility in controlling the number of AJSCC levels and also achieves better performance compared with the first design. The presented biosensor system has been evaluated via both hardware experiments as well as wireless link simulations. Future work will include developing a classifier on the digital receiver to classify the cells based on impedance pulse signals so as to estimate multiple biomarker concentrations.

\section*{Acknowledgments} 
We thank Niloy Talukder for designing the lock-in amplifier in the microfluidic system, and Ajinkya Padwad, Karan Ahuja, Nikitha Maiya and Anthony Yang for helping in experiments.

\bibliographystyle{IEEEtran}
\bibliography{ref_cytometry_ajscc}

\begin{thebibliography}{10}
\providecommand{\url}[1]{#1}
\csname url@samestyle\endcsname
\providecommand{\newblock}{\relax}
\providecommand{\bibinfo}[2]{#2}
\providecommand{\BIBentrySTDinterwordspacing}{\spaceskip=0pt\relax}
\providecommand{\BIBentryALTinterwordstretchfactor}{4}
\providecommand{\BIBentryALTinterwordspacing}{\spaceskip=\fontdimen2\font plus
\BIBentryALTinterwordstretchfactor\fontdimen3\font minus
  \fontdimen4\font\relax}
\providecommand{\BIBforeignlanguage}[2]{{%
\expandafter\ifx\csname l@#1\endcsname\relax
\typeout{** WARNING: IEEEtran.bst: No hyphenation pattern has been}%
\typeout{** loaded for the language `#1'. Using the pattern for}%
\typeout{** the default language instead.}%
\else
\language=\csname l@#1\endcsname
\fi
#2}}
\providecommand{\BIBdecl}{\relax}
\BIBdecl

\bibitem{Zhao17}
X.~Zhao, V.~Sadhu, T.~Le, M.~Javanmard, and D.~Pompili, ``Towards low-power
  wearable wireless sensors for molecular biomarker and physiological signal
  monitoring,'' in \emph{IEEE International Symposium on Circuits and Systems
  (ISCAS)}, May 2017.

\bibitem{Soh15}
P.~J. Soh, G.~A.~E. Vandenbosch, M.~Mercuri, and D.~M. M.~P. Schreurs,
  ``Wearable wireless health monitoring: Current developments, challenges, and
  future trends,'' \emph{IEEE Microwave Magazine}, vol.~16, no.~4, pp. 55--70,
  May 2015.

\bibitem{Rajendra17}
V.~Rajendra and O.~Dehzangi, ``Detection of distraction under naturalistic
  driving using galvanic skin responses,'' in \emph{2017 IEEE 14th
  International Conference on Wearable and Implantable Body Sensor Networks
  (BSN)}, May 2017, pp. 157--160.

\bibitem{Yeo16}
\BIBentryALTinterwordspacing
J.~C. Yeo, Kenry, and C.~T. Lim, ``Emergence of microfluidic wearable
  technologies,'' \emph{Lab Chip}, vol.~16, pp. 4082--4090, 2016. [Online].
  Available: \url{http://dx.doi.org/10.1039/C6LC00926C}
\BIBentrySTDinterwordspacing

\bibitem{Sun16}
A.~Sun, A.~G. Venkatesh, and D.~A. Hall, ``A multi-technique reconfigurable
  electrochemical biosensor: Enabling personal health monitoring in mobile
  devices,'' \emph{IEEE Transactions on Biomedical Circuits and Systems},
  vol.~10, no.~5, pp. 945--954, Oct 2016.

\bibitem{EJ12}
\BIBentryALTinterwordspacing
S.~Emaminejad, M.~Javanmard, R.~W. Dutton, and R.~W. Davis, ``Microfluidic
  diagnostic tool for the developing world: contactless impedance flow
  cytometry,'' \emph{Lab Chip}, vol.~12, pp. 4499--4507, 2012. [Online].
  Available: \url{http://dx.doi.org/10.1039/C2LC40759K}
\BIBentrySTDinterwordspacing

\bibitem{GasterHall09}
R.~S. Gaster, D.~A. Hall, C.~H. Nielsen, S.~J. Osterfeld, H.~Yu, K.~E. Mach,
  R.~J. Wilson, B.~Murmann, J.~C. Liao, S.~S. Gambhir, and S.~X. Wang,
  ``Matrix-insensitive protein assays push the limits of biosensors in
  medicine,'' \emph{Nat Med}, vol.~15, no.~11, pp. 1327--1332, Nov 2009.

\bibitem{GaoEmaminejad16}
W.~Gao, S.~Emaminejad, H.~Y.~Y. Nyein, S.~Challa, K.~Chen, A.~Peck, H.~M.
  Fahad, H.~Ota, H.~Shiraki, D.~Kiriya, D.-H. Lien, G.~A. Brooks, R.~W. Davis,
  and A.~Javey, ``Fully integrated wearable sensor arrays for multiplexed in
  situ perspiration analysis,'' \emph{Nature}, vol. 529, no. 7587, pp.
  509--514, Jan 2016, letter.

\bibitem{Gholizadeh17}
A.~Gholizadeh, D.~Voiry, C.~Weisel, A.~Gow, R.~Laumbach, H.~Kipen,
  M.~Chhowalla, and M.~Javanmard, ``Toward point-of-care management of chronic
  respiratory conditions: Electrochemical sensing of nitrite content in exhaled
  breath condensate using reduced graphene oxide,'' vol.~3, pp. 17\,022 EP --,
  May 2017, article.

\bibitem{Lin15}
Z.~Lin, X.~Cao, P.~Xie, M.~Liu, and M.~Javanmard, ``Picomolar level detection
  of protein biomarkers based on electronic sizing of bead aggregates:
  theoretical and experimental considerations,'' \emph{Biomedical
  Microdevices}, vol.~17, no.~6, p. 119, 2015.

\bibitem{Mok14}
J.~Mok, M.~N. Mindrinos, R.~W. Davis, and M.~Javanmard, ``Digital microfluidic
  assay for protein detection,'' \emph{Proceedings of the National Academy of
  Sciences}, vol. 111, no.~6, pp. 2110--2115, 2014.

\bibitem{Javanmard11}
M.~Javanmard and R.~Davis, ``A microfluidic platform for electrical detection
  of dna hybridization,'' \emph{Sensors and Actuators B: Chemical}, vol. 154,
  no.~1, pp. 22 -- 27, 2011, transducers 2009.

\bibitem{Talukder17}
N.~Talukder, A.~Furniturewalla, T.~Le, M.~Chan, S.~Hirday, X.~Cao, P.~Xie,
  Z.~Lin, A.~Gholizadeh, S.~Orbine, and M.~Javanmard, ``A portable battery
  powered microfluidic impedance cytometer with smartphone readout: towards
  personal health monitoring,'' \emph{Biomedical Microdevices}, vol.~19, no.~2,
  p.~36, Apr 2017.

\bibitem{Kario17}
K.~Kario, N.~Tomitani, H.~Kanegae, and et.al., ``Development of a new ict-based
  multisensor blood pressure monitoring system for use in hemodynamic
  biomarker-initiated anticipation medicine for cardiovascular disease: The
  national impact program project,'' \emph{Progress in Cardiovascular
  Diseases}, Dec 2017.

\bibitem{Hu16}
J.~Hu, X.~Cui, Y.~Gong, X.~Xu, B.~Gao, T.~Wen, T.~J. Lu, and F.~Xu, ``Portable
  microfluidic and smartphone-based devices for monitoring of cardiovascular
  diseases at the point of care,'' \emph{Biotechnology Advances}, vol.~34,
  no.~3, pp. 305 -- 320, 2016, trends in In Vitro Diagnostics and Mobile
  Healthcare.

\bibitem{Imani16}
S.~Imani, A.~J. Bandodkar, A.~M.~V. Mohan, R.~Kumar, S.~Yu, J.~Wang, and P.~P.
  Mercier, ``A wearable chemical-lectrophysiological hybrid biosensing system
  for real-time health and fitness monitoring,'' \emph{Nature Communications},
  vol.~7, pp. 11\,650 EP --, May 2016, article.

\bibitem{Luhmann16}
A.~von Luhmann, H.~Wabnitz, T.~Sander, and K.~R. Muller, ``M3ba: A mobile,
  modular, multimodal biosignal acquisition architecture for miniaturized
  eeg-nirs based hybrid bci and monitoring,'' \emph{IEEE Transactions on
  Biomedical Engineering}, vol.~PP, no.~99, pp. 1--1, 2016.

\bibitem{Abrar16}
M.~A. Abrar, Y.~Dong, P.~K. Lee, and W.~S. Kim, ``Bendable electro-chemical
  lactate sensor printed with silver nano-particles,'' \emph{Scientific
  Reports}, vol.~6, pp. 30\,565 EP --, Jul 2016, article.

\bibitem{Nemiroski14}
A.~Nemiroski, D.~C. Christodouleas, J.~W. Hennek, A.~A. Kumar, E.~J. Maxwell,
  M.~T. Fern{\'a}ndez-Abedul, and G.~M. Whitesides, ``Universal mobile
  electrochemical detector designed for use in resource-limited applications,''
  \emph{Proceedings of the National Academy of Sciences}, vol. 111, no.~33, pp.
  11\,984--11\,989, 2014.

\bibitem{SensorsJournal18}
X.~Zhao, V.~Sadhu, A.~Yang, and D.~Pompili, ``Improved circuit design of analog
  joint source channel coding for low-power and low-complexity wireless
  sensors,'' \emph{IEEE Sensors Journal}, vol.~18, no.~1, pp. 281--289, Jan
  2018.

\bibitem{Zhao16}
X.~Zhao, V.~Sadhu, and D.~Pompili, ``Low-power all-analog circuit for
  rectangular-type analog joint source channel coding,'' in \emph{2016 IEEE
  International Symposium on Circuits and Systems (ISCAS)}, May 2016.

\bibitem{wons3tier2017}
V.~Sadhu, X.~Zhao, and D.~Pompili, ``Energy-efficient analog sensing for
  large-scale, high-density persistent wireless monitoring,'' in \emph{IEEE
  13th Annual Conference on Wireless On-demand Network Systems and Services
  (WONS)}, Feb 2017, pp. 1--8.

\bibitem{MASS17}
X.~Zhao, V.~Sadhu, and D.~Pompili, ``Analog signal compression and multiplexing
  techniques for healthcare internet of things,'' in \emph{2017 IEEE 14th
  International Conference on Mobile Ad Hoc and Sensor Systems (MASS)}, Oct
  2017, pp. 398--406.

\bibitem{Emaminejad12}
S.~Emaminejad, M.~Javanmard, R.~W. Dutton, and R.~W. Davis, ``Microfluidic
  diagnostic tool for the developing world: contactless impedance flow
  cytometry,'' \emph{Lab Chip}, vol.~12, pp. 4499--4507, 2012.

\bibitem{balakrishnan2013node}
K.~R. Balakrishnan, G.~Anwar, M.~R. Chapman, T.~Nguyen, A.~Kesavaraju, and
  L.~L. Sohn, ``Node-pore sensing: a robust, high-dynamic range method for
  detecting biological species,'' \emph{Lab on a Chip}, vol.~13, no.~7, pp.
  1302--1307, 2013.

\bibitem{le2016biomems}
T.~Le, G.~Salles-Loustau, L.~Najafizadeh, M.~Javanmard, and S.~Zonouz,
  ``Biomems-based coding for secure medical diagnostic devices,'' in
  \emph{Engineering in Medicine and Biology Society (EMBC), 2016 IEEE 38th
  Annual International Conference of the}.\hskip 1em plus 0.5em minus
  0.4em\relax IEEE, 2016, pp. 4419--4422.

\bibitem{xia1998soft}
Y.~Xia and G.~M. Whitesides, ``Soft lithography,'' \emph{Annual review of
  materials science}, vol.~28, no.~1, pp. 153--184, 1998.

\bibitem{duffy1998rapid}
D.~C. Duffy, J.~C. McDonald, O.~J. Schueller, and G.~M. Whitesides, ``Rapid
  prototyping of microfluidic systems in poly (dimethylsiloxane),''
  \emph{Analytical chemistry}, vol.~70, no.~23, pp. 4974--4984, 1998.

\bibitem{tan2010oxygen}
S.~H. Tan, N.-T. Nguyen, Y.~C. Chua, and T.~G. Kang, ``Oxygen plasma treatment
  for reducing hydrophobicity of a sealed polydimethylsiloxane microchannel,''
  \emph{Biomicrofluidics}, vol.~4, no.~3, p. 032204, 2010.

\bibitem{opamp}
W.~S. Y.~Libin, M.~Steyaert, ``A 0.8{V} 8 $\mu${W} {CMOS OTA} with 50-d{B} gain
  and 1.2-{MHz} {GBW} in 18-p{F} load,'' Jun. 2003, pp. 297--300.

\bibitem{comparator}
A.~Valaee and M.~Maymandi-Nejad, ``An ultra low-power low-voltage track and
  latch comparator,'' in \emph{Proc. of IEEE International Conference on
  Electronics, Circuits, and Systems (ICECS)}, Dec. 2010, pp. 186--189.

\bibitem{discrete_ic}
``{Sizing Up Discrete Devices Against Integrated Circuits},''
  \url{http://www.mwrf.com/analog-semiconductors/sizing-discrete-devices-against-integrated-circuits}.

\bibitem{wsn430}
``{WSN430 Open Node},'' \url{https://www.iot-lab.info/hardware/wsn430/}.

\bibitem{telosb}
``{MoteIV Telos (RevB) Low Power Wireless Sensor Module},''
  \url{http://www.memsic.com/userfiles/files/Datasheets/WSN/telosb\_datasheet.pdf}.

\bibitem{Zhao07}
X.~Zhao, R.~S.~K. Cheng, and D.~C.~Y. Ong, ``A comparative analysis of pilot
  placement schemes in frequency-selective fast fading mimo channel,'' in
  \emph{2007 Wireless Telecommunications Symposium}, April 2007, pp. 1--7.

\bibitem{Sarnoff07}
X.~Zhao, D.~C.~K. Lee, Z.~Pan, N.~Boubaker, and R.~S.~K. Cheng, ``Modified
  antenna effective gain in multiple-cluster 3d channel model,'' in \emph{2007
  IEEE Sarnoff Symposium}, April 2007, pp. 1--5.

\bibitem{Ali94}
M.~H. Ali, A.~S. Parker, and K.~Pahlavan, ``Frequency domain model for standard
  simulation of wideband radio propagation for personal communications,''
  \emph{Electronics Letters}, vol.~30, no.~25, pp. 2103--2104, Dec 1994.

\end{thebibliography}

\begin{IEEEbiography}[{\includegraphics[width=1in,height=1.25in,clip,keepaspectratio]{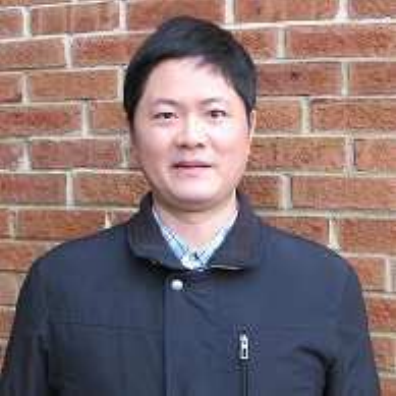}}]{Xueyuan~Zhao}
received the bachelor’s and master’s degrees from the Beijing University of Posts and Telecommunications (BUPT). He is currently pursuing the Ph.D. degree with the Electrical and Computer Engineering Department, Rutgers University, New Brunswick. He is a member of the Cyber-Physical Systems Laboratory directed by Dr. D. Pompili since 2013. He worked in industry after graduation from BUPT, and contributed to multiple research and development projects. He is an inventor and co-inventor of a number of U.S. patents. He was a co-recipient of the Qualcomm Innovation Fellowship Finalist Award.
\end{IEEEbiography}

\begin{IEEEbiography}[{\includegraphics[width=1in,height=1.25in,clip,keepaspectratio]{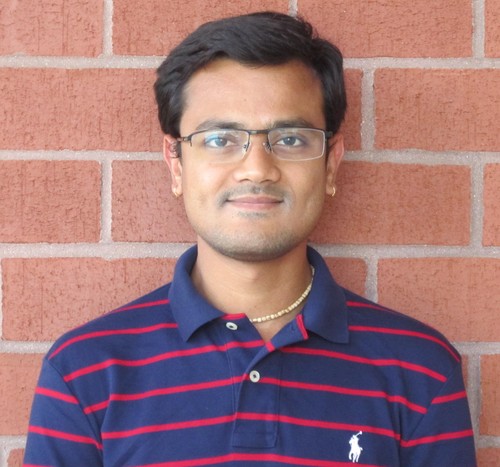}}]{Vidyasagar~Sadhu} (S’17) received the B.Tech. and M.Tech. degrees (under dual degree program) in ECE from the Indian Institute of Technology, Chennai, India, in 2012. He worked in industry for two years before joining the Ph.D. Program in Cyber Physical Systems Laboratory led by Dr. D. Pompili with the Department of Computer Engineering, Rutgers University, in 2014. His research interests are in the domain of distributed computing and machine learning, mobile phone sensing, planning under uncertainty and wireless sensor networks. He has contributed to several research projects, a co-recipient of the NSF Student Research Grant Award and the Qualcomm Innovation Fellowship Finalist Award.
\end{IEEEbiography}

\begin{IEEEbiography}[{\includegraphics[width=1in,height=1.25in,clip,keepaspectratio]{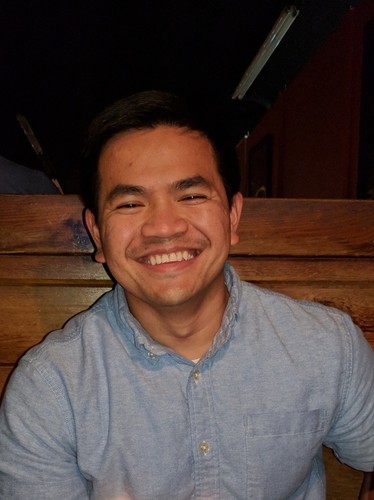}}]{Tuan~Le} received the B.Sc. degree in electrical and computer engineering from Rutgers University in 2012, where he is currently pursuing the Ph.D. degree with NanoBioelectronics Laboratory. He joined Teletronics Technology Corporation for the duration of one and a half years as Support Engineer. His work function included RF radio system for high-speed communication and aircraft test instrumentation. His research interests include biosensors and security for point-of- care diagnostic devices.
\end{IEEEbiography}

\begin{IEEEbiography}[{\includegraphics[width=1in,height=1.25in,clip,keepaspectratio]{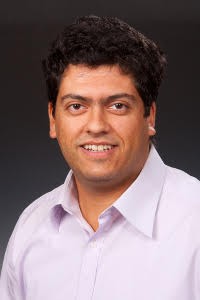}}]{Dario~Pompili} received the Laurea (combined B.S. and M.S.) and Doctorate degrees in telecommunications and system engineering from the University of Rome La Sapienza, Italy, in 2001 and 2004, respectively, and the Ph.D. degree in ECE from the Georgia Institute of Technology in 2007. He is currently an Associate Professor with the Department of ECE, Rutgers University. He is the Director of the Cyber-Physical Systems Laboratory, which focuses on mobile computing, wireless communications and networking, acoustic communications, sensor networks, and datacenter management. He was a recipient of the NSF CAREER’11, ONR Young Investigator Program’12, and DARPA Young Faculty’12 awards. In 2015, he was a nominated Rutgers–New Brunswick Chancellor’s Scholar. He has published more than 100 refereed scholar publications: with over 7,000 citations, he has an h-index of 31 and an i10-index of 63 (Google Scholar, in April 2018). He is a Senior Member of the IEEE Communications Society and the ACM.
\end{IEEEbiography}

\begin{IEEEbiography}[{\includegraphics[width=1in,height=1.25in,clip,keepaspectratio]{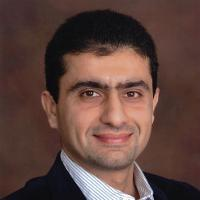}}]{Mehdi~Javanmard} joined Electrical and Computer Engineering Department at Rutgers University in Fall 2014 as Assistant Professor. Before that he was Senior Research Engineer at the Stanford Genome Technology Center (SGTC) in the School of Medicine at Stanford University. He received his BS (2002) from Georgia Institute of Technology and the MS in Electrical Engineering at Stanford University (2004) working at Stanford Linear Accelerator Center researching the use of photonic nanostructures for high energy physics. In 2008, he received his PhD in Electrical Engineering at Stanford University working on development of electronic microfluidic platforms for low cost genomic and proteomic biomarker detection. At SGTC, he worked as a postdoctoral scholar from 2008-2009, and then as a staff engineering research associate from 2009 till 2014. In 2017 he was recipient of the Translational Medicine and Therapeutics Award by the American Society for Clinical Pharmacology \& Therapeutics for his group's work in point of care diagnostic tools for assessing patient response to cancer therapies. He has received various awards as Principal Investigator from the National Science Foundation, DARPA, and the PhRMA foundation to support his research. His interests lie in developing portable and wearable technologies for continuous health monitoring and understanding the effects of environment on health.
\end{IEEEbiography}

\end{document}